TECHNICAL   REPORT   # 03 – 1

# Temperature Dependence of Microwave Loss, Electronic Transport and Magnetic Parameters in Polycrystalline Perovskite Manganites – Evidence that the Magnetic and the Metal-Insulator Transitions are not Identical

S. I. Patil and S. M. Bhagat, FMR Group, Dept. of Physics, University of Maryland, College Park MD – 20742

Kristy Enckson, Chris Lanci and S. E. Lofland, Dept. of  Physics, Rowan University, Glassboro, NJ 08028

M. S. Sahasrabudhe and S. N. Sadakale, Dept. of Physics, University of Pune- 410007, India

Abstract

Measurements of the variation with temperature of the dc resistivity, ac susceptibility, dc magnetization and microwave loss for several systems of pseudo-cubic manganites reveal that the characteristic temperatures delineated by the various properties can be very The observations further suggest that both the transitions are percolative in nature and therefore there is no a priori reason for them to be identical.

Introduction:

In perovskite manganites of the form $R_{1-y}A_yMnO_3$, where R is a trivalent rare earth and A a divalent alkali earth, the spin, lattice, charge and orbital degrees of freedom are coupled to one another. Since the interaction energies are also of the same order of magnitude their properties are extremely sensitive to small changes in the material parameters thereby leading to a very rich phase diagram.[1] In early studies[2] it was found that for $y = 0.3$, for example, there was a paramagnetic (PM) to ferromagnetic (FM) transition at the Curie temperature $T_C$ and at roughly the same temperature a high temperature "insulating" (I) phase underwent a transition to a "metallic" (M) phase as marked by a peak in the resistivity versus temperature diagram, hence $T_p$. This motivated the introduction of the Zener double exchange mechanism[3] wherein the onset of FM coupling facilitated transport of the holes in the $Mn^{4+}$-$Mn^{3+}$ complex thereby leading to a drop in resistivity $\rho$.

In a variety of situations $T_C$ and $T_p$ appeared to be quite close to one another and this has led to the lore that they are invariably the same and that the PM-FM and I-M transitions are concomitant. In assessing such claims one must keep in mind that extensive ferromagnetic resonance studies[4] of doped manganites reveal that nearly all the samples are magnetically quite inhomogeneous. Also, the "transitions" are far from sharp and that, more often than not, there is a distribution of $T_C$ values in a sample.[5] On the other hand, it is notable that in a careful investigation[6] of highly well characterized single crystal samples of $La_{0.8}Sr_{0.2}MnO_3$ it has been found that $T_p$ is nearly 20 K higher than $T_C$ and since the drop in $\rho$ is rapid, a sizable part of the I–M transition is complete well before any bulk FM phase appears.

Recent theoretical developments[7] also suggest that the I–M and PM–FM transitions need not be identical. Such findings have motivated the present studies, wherein we have prepared a wide variety of doped manganites of both the wide-band ($La_{1-y}Sr_yMnO_3$) and narrow-band ($La_{1-y}Ca_yMnO_3$) types and studied the temperature dependencies of electronic transport as well as the magnetic transitions. It turns out that not only $T_p$ and $T_C$ differ from one another, but also in some cases one of the transitions is hysteretic in temperature while the other shows no hysteresis. Supplementing these measurements with microwave loss studies reveals the percolative nature of some of the I–M transformations. Thus, there are both qualitative and quantitative distinctions between the I–M and PM–FM transformations. This should further motivate theoretical investigations that take explicit account of the disorder in these materials. For the present, it seems reasonable to speculate that there exist two types of percolative behavior – magnetic and conductive and there is no a priori reason for them to have the same onset temperature.

Experimental Methods:

Polycrystalline samples of the series $(La_{1-x}Pr_x)_{0.7}Sr_{0.3}MnO_3$ ($x$ = 0, 0.2, 0.3, 0.4, 0.5, 0.6, 0.8, 1), $(La_{1-x}Pr_x)_{0.7}Ca_{0.3}MnO_3$ ($x$ = 0, 0.2, 0.4, 0.5, 0.6, 0.7 0.8) and $(La_{1-y}Ba_yMnO_3$ ($y$ =0.5, 0.51, 0.55, 0.58, 0.60, 0.65) were synthesized by the solid-state reaction method starting from the respective oxides. The single-phase nature of the samples was established by x-ray diffraction. The electrical transport measurements were carried out by four-probe method for temperature $T$ ranging between 380 K and 5 K. The magnetic transition temperature was measured with a low-field (12 Oe) homemade ac susceptometer operating at 135 Hz with 80<$T$<380 K. As described earlier,[8] in this instrument one measures the ac susceptibility $\chi_{ac}$ over the central 2 mm of a 1-cm long parallelepiped sample. Microwave losses for 77 < $T$ < 360 K were measured by a conventional spectrometer[9] operating at 9.87GHz with the cavity perturbation technique. The magnetization measurements were carried out with dc extraction and SQUID magnetometers made by Quantum Design.

Results:

**1 AC Susceptibility**

A : $(La_{1-x}Pr_x)_{0.7}Sr_{0.3}MnO_3$

Figure 1a presents the $\chi_{ac}$ data for the $(La_{1-x}Pr_x)_{0.7}Sr_{0.3}MnO_3$ series for $x$ = 0, 0.2, 0.3, 0.4, 0.5, 0.6, and Fig.1b shows the $\chi_{ac}$ for $x$ = 0.8 and 1 as a function of temperature. In every case one observes a "step" in $\chi_{ac}$, which marks the PM – FM transition. As before,[8] it is noted that in these samples $\chi_{ac}$ is demagnetization limited and hence constant at low temperature. Thus the PM–FM transition can be said to be "complete." In good agreement with previous data,[1] the $x$ = 0 sample has a very sharp transition at 361 K. $T_C$ decreases with an increase in $x$. It is found that the transition is sharp for $x$ < 0.6 and there is no thermal hysteresis. Thermal hysteresis of ≥ 10 K starts developing with $x$ > 0.6 and the transition becomes broader at higher values of $x$. The $x$ = 0.8 and 1 samples show clear thermal hysteresis.

B: $(La_{1-x}Pr_x)_{0.7}Ca_{0.3}MnO_3$

Figure 2 shows $\chi_{ac}$ as a function of temperature for the system $(La_{1-x}Pr_x)_{0.7}Ca_{0.3}MnO_3$ with $x$ = 0, 0.2, 0.4, 0.5, 0.6, 0.7 respectively. For higher values of $x$, $\chi_{ac}$ fails to "saturate" on cooling for nearly 40 K below the initial rise, suggesting that the FM phase is not pervading the entire sample. It is notable that in every case there is thermal hysteresis; that is, the cooling and heating curves do not coincide.[10] Presumably, for $x > 0.5$ the hysteresis appears to be less marked because as noted above the FM phase is not fully established over the available low temperature range. The gradual onset of bulk FM points to a percolative origin for the PM–FM transition.

C : $La_{1-y}Ba_yMnO_3$:

Previously,[11] we have studied the $La_{0.67}Ba_{0.33}MnO_3$ system and it was found that the $\chi_{ac}$ transition at $T_C$ is very sharp at 343 K as shown in Fig. 3a. In addition it appeared that $T_p$ and $T_C$ values were very close to each other. However, FMR data[5] clearly pointed to a sizeable (10 K) distribution of $T_C$ values in many of the samples. To investigate the dependence of $T_C$ measured by $\chi_{ac}$ on the values of $y$, we have synthesized and studied $La_{1-y}Ba_yMnO_3$ compounds with $0.5 < y < 0.65$. Figure 3b shows the variation $\chi_{ac}$ with temperature for $y = 0.50, 0.51, 0.55, 0.58, 0.60, 0.65$. All the samples show PM–FM transition located at $\sim 330 < T_C < 340K$. These transitions are not as sharp as in $y = 0.33$, but they are much narrower than the transitions in Figs. 1 and 2.

**2 dc Resistance**

A : $(La_{1-x}Pr_x)_{0.7}Sr_{0.3}MnO_3$

Figure 4a shows the resistance $R_T$, normalized to its value at 5 K for $(La_{0.2}Pr_{0.8})_{0.7}Sr_{0.3}MnO_3$ sample, and it is clear that material has a sharp I-M transition at 295 K. In the metallic region ($5 < T < T_p - 30$ K) $R_T$ follows a $T^2$ dependence as shown in Fig 4b. The slope of the curve is found to be $5 \times 10^{-5}$ K$^{-2}$, well in accord with previous observations.[1,12] It is arguable that it arises from magnon scattering. Figure 4c shows the normalized resistance vs. temperature plots for all other values of $x$ for $(La_{1-x}Pr_x)_{0.7}Sr_{0.3}MnO_3$ series. The $T_p$ value decreases with increase in $x$, and interestingly it is observed that there is no thermal hysteresis between cooling and heating cycles even though the magnetic transitions show (Fig.1b) thermal hysteresis in $x = 0.8$ and 1. Again, over a wide temperature ranges ($5 < T < T_p -30$ K), the normalized resistance follows a $T^2$ dependence as shown in Fig. 4d, and all the slopes are of the same order of magnitude ($\sim 10^{-5}$ K$^{-2}$). As before,[1] the $x = 0$ and 0.2 samples (not shown) did not exhibit any metal-insulator transition. Rather, the resistance decreases continuously with decrease in temperature.

B : $(La_{1-x}Pr_x)_{0.7}Ca_{0.3}$ MnO$_3$

Fig.5a shows a typical plot of $R_T$ (normalized at 300 K) as a function of $T$ for $x = 0.4$ sample while Fig.5b display the $T$ dependence of $R_T$ different values of $x$. As $x$ increases, $T_p$ shifts lower[10] and thermal hysteresis starts developing with $x > 0.5$. $R_T$ at $T_p$ increases with $x$, and also the drop at $T < T_p$ becomes sharper. Further $R_T(T_p)$ is found to be higher during cooling as compared the value during heating. Whereas $R_T(T_p)$ increases slowly as $x$ increases from 0 to 0.6, there is a sharp jump as $x$ becomes larger. Finally, for $x = 1$, (not shown) an insulating charge ordered state appears. Even though, the ac susceptibility (Fig.2c) shows thermal hysteresis for $x = 0.4$ sample, $R_T$ does not show any thermal hysteresis in the sense that $T_p$ is the same during both heating and cooling even though $R_T$ may be different.

C : $La_{1-y}Ba_yMnO_3$

The temperature dependence of $R_T$ (normalized at 300 K) for $y = 0.6$ is shown in Fig. 6. The I–M transition is not at all sharp, but it is clear that the electron transport changes its character for $T < T_p$. It is not possible to mark $T_p$ to better than $\pm 5$ K.

## 3 Zero-field Microwave Absorption

A : $(La_{1-x}Pr_x)_{0.7}Sr_{0.3}MnO_3$

Figure 7a shows the temperature dependence of the zero-field microwave absorption in $(La_{0.2}Pr_{0.8})_{0.7}Sr_{0.3}MnO_3$ in the temperature range just below its $T_C$, and one notes a sharp minimum. As reported in previous studies,[13] this dip arises because in this material the surface impedance is largely controlled by the dynamic permeability rather than by the resistivity. In bulk ferromagnetic samples this will cause a minimum in the absorption when the magnetization $4\pi M$ becomes equal to $\omega/\gamma$ ( = 3.5 kOe for the present frequency with $g = 2$), where $\omega$ is the angular frequency and $\gamma = g\mu_B/\hbar$ the gyromagnetic ratio. Thus, this is a case of the temperature-tuned ferromagnetic antiresonance (FMAR) phenomena discussed in detail.[13] In fact as seen in Fig. 7b, where we have collected together the observed temperature dependences for all the $(La_{1-x}Pr_x)_{0.7}Sr_{0.3}MnO_3$ samples, in every case we observed only the FMAR effect rather than the peak which would be anticipated from the $T$ dependence of $R_T$ shown in Fig. 4c. In addition it is notable that for $x = 0.8$ and 1, the temperature of the dip during cooling is several degrees lower than while heating. That is, the material has the same $M$ value at two different temperatures; a clear indication that the magnetic transition has first-order character.

B : $(La_{1-x}Pr_x)_{0.7}Ca_{0.3}MnO_3$

The $T$ dependence of the microwave loss for $x = 0.2$ and $x = 0.4$ is shown in Figs. 8 a and b, respectively. In both cases, the thermal hysteresis is obvious although the resistance (Fig. 5) shows no such difference between cooling and heating cycles. In addition, as noted previously, the peak in microwave absorption occurs at a temperature well below $T_p$. That is, although $R_T$ is dropping for $T_C < T_p$, the microwave loss keeps increasing. As discussed below, this is a symptom of the percolative nature of the I–M transition.

The $T$ dependence of the microwave loss for $x = 0.6, 0.7, 0.8$ as shown in Fig. 9 is more complex. In these materials the dc resistivity is much higher and therefore the finite size effect[14] comes into play. For example, in $x = 0.6$ (0.7), with a 1-mm thick sample, the loss is "inverted" for temperatures lying between 225 K (240 K) and 160 K (130 K) during cooling. That is the microwave loss is controlled by $1/\rho^n_{dc}$, with $n$ varying between 0.5 and 1, rather than $\rho_{dc}^{1/2}$, and therefore the absorption exhibits a broad minimum around $T_p$. The low-$T$ peak (at 160 K in 0.6 and 130 K in 0.7) marks the return to normal ($\rho_{dc}^{1/2}$) dependence at lower temperature and has no fundamental significance. To check this further a 0.3-mm thick sample of $x = 0.7$ was studied (Fig. 10). As expected the low-$T$ peak shifted to lower temperatures and also at the high-$T$ end the inversion temperature moved to > 300 K. The existence of this finite size effect is somewhat unfortunate as it prevents one from studying the percolative nature of the I-M transition at higher $x$ with the microwave technique.

## 4. dc Magnetization

A : $(La_{1-x}Pr_x)_{0.7}Sr_{0.3}MnO_3$

The low temperature magnetic isothermals show that for all $x$ the spontaneous magnetization is around 90 – 95 emu/g (Table I). That is, we are observing the full Mn moment. The 5-kOe isotherms for $x \geq 0.4$, shown in Fig. 11, exhibit a slight thermal hysteresis as expected. By analogy with the liquid–gas case, $H$ here plays the role of pressure so we should anticipate that at higher $H$ the cooling and heating curves coincide.

B : $(La_{1-x}Pr_x)_{0.7}Ca_{0.3}MnO_3$

The low temperature magnetic isothermals show that for all $x$ the spontaneous magnetization is around 95 – 100 emu/g (Table II). That is, again we are observing the full Mn moment. The saturated moment is excessively large, up to 120 emu/g. Presumably, this is partially an effect of the Pr, although this is not observed in $(La_{1-x}Pr_x)_{0.7}Sr_{0.3}MnO_3$. Again, the 5-kOe isotherms for $x \geq 0.4$, shown in Fig. 12, exhibit thermal hysteresis as expected. Field sweeps at 5 K for $x < 0.8$ show simple behavior as expected for a ferromagnet (Fig. 13 a). However, for larger $x$, there are steps in the magnetization which appear in the virgin curve but not on subsequent sweeps. For $x = 1$ (Fig. 13 d), there are many steps. This is similar to what has been found on similar compounds.[15]

Discussion

The apparent transition temperatures for the three sets of samples are collected together in Fig. 14. In the situation when the $\chi_{ac}$ vs $T$ curves are different for heating and cooling, one can no longer talk of a critical temperature in the strict sense of the term. By analogy with the liquid-gas case as one can speak of a "boiling" temperature. That said, we label the respective temperatures $T_C^C$ and $T_C^H$ for cooling and heating respectively. It is immediately clear that $T_p$ and $T_C$ can be significantly different from one another. Amazingly, whereas the PM–FM transition follows (in the sense of lowering $T$) the I–M change in $(La_{1-x}Pr_x)_{0.7}Sr_{0.3}MnO_3$ and $(La_{1-x}Pr_x)_{0.7}Ca_{0.3}MnO_3$, the reverse is true for $La_{1-y}Ba_yMnO_3$, the discrepancy rising to near the 100 K at $x \sim 0.6$. In addition, while the magnetic transitions show significant thermal hysteresis, the corresponding resistive transition has no hysteresis. That is, the PM-FM transition is of first order but not so the

I-M transformation. Note also that when the PM-FM transition is hysteretic, there is no anomaly in $\rho$ in that temperature region making it unlikely that a structural transition is occurring. The non-coincidence of the peak in the microwave loss and $T_p$, in $(La_{0.8}Pr_{0.2})_{0.7}Ca_{0.3}MnO_3$ and $(La_{0.6}Pr_{0.4})_{0.7}Ca_{0.3}MnO_3$ is best understood by realizing that whereas $\rho_{dc}$ can drop on establishment of a highly conducting percolating path, the microwave loss will reduce only when the entire sample becomes less resistive.

It seems reasonable to conclude that the two transitions in the mixed manganites are both percolative in nature, and there is a no prior reason to claim that one is concomitant with the other.

**Acknowledgments**

This work was supported in part by NSF MRSEC 00-80008 and the New Jersey Commission on Higher Education.

Figure Captions

Fig. 1 Temperature dependence of the ac susceptibility of for a) $x < 0.6$ and b) $x = 0.8$ and 1. The latter shows thermal hysteresis.

Fig. 2 Temperature dependence of the ac susceptibility of $(La_{1-x}Pr_x)_{0.7}Ca_{0.3}MnO_3$ for a) $x =0$, b) $x = 0.2$, c) $x = 0.4$, d) $x = 0.5$, e) $x = 0.6$ and f) $x = 0.7$.

Fig. 3 Temperature dependence of the ac susceptibility of $La_{1-y}Ba_yMnO_3$ for a) $y=0.33$, b) $y=0.50, 0.51, 0.55, 0.58, 0.60$, and $0.65$.

Fig. 4 Dependence of the normalized resistivity of $(La_{0.2}Pr_{0.8})_{0.7}Sr_{0.3}MnO_3$ on a) temperature and b) temperature squared. Dependence of the normalized resistivity of $(La_{1-x}Pr_x)_{0.7}Sr_{0.3}MnO_3$ on d) temperature and d) temperature squared. The $T^2$ dependence could be attributed electron-magnon scattering.

Fig. 5 Temperature dependence of the normalized resistivity of a) $(La_{0.6}Pr_{0.4})_{0.7}Ca_{0.3}MnO_3$ and b) $(La_{1-x}Pr_x)_{0.7}Ca_{0.3}MnO_3$. Hysteresis is observed only for $x \geq 0.5$.

Fig. 6 Temperature dependence of the normalized resistivity $La_{0.4}Ba_{0.6}MnO$.

Fig. 7 Microwave absorption of $(La_{1-x}Pr_x)_{0.7}Sr_{0.3}MnO_3$ for a) $x=0.8$, b) all $x$. Due to the metallic nature of the resistivity, the absorption is controlled by the dynamic permeability and thus the magnetization. For $x \geq 0.8$, there is thermal hysteresis.

Fig. 8 Microwave absorption of $(La_{1-x}Pr_x)_{0.7}Ca_{0.3}MnO_3$ for a) $x=0.2$, b) $x=0.4$. Here the losses are dominated by the electronic transport, but he microwave losses drop at a temperature well below $T_p$.

Fig. 9 Microwave absorption of $(La_{1-x}Pr_x)_{0.7}Ca_{0.3}MnO_3$ for a) $x=0.6$, b) $x=0.7$, and c) $x=0.8$, all of which show thermal hysteresis. Due to the nearly insulating behavior, the absorption is complex function.

Fig. 10 Microwave absorption of two samples of $(La_{0.3}Pr_{0.7})_{0.7}Ca_{0.3}MnO_3$, demonstrating the effect of size on the microwave absorption in highly resistive compounds.

Fig. 11 Temperature dependence of the magnetization of $(La_{1-x}Pr_x)_{0.7}Sr_{0.3}MnO_3$ for a) $x=0.2$, b) $x=0.5$, c) $x=0.6$, and d) $x=0.8$. For larger values $x$, one begins to see thermal hysteresis.

Fig. 12 Temperature dependence of the magnetization of $(La_{1-x}Pr_x)_{0.7}Ca_{0.3}MnO_3$ for a) $x=0.2$, b) $x=0.4$, c) $x=0.6$, d) $x=0.7$, e) $x=0.8$, f) $x=0.9$, and g) $x=1$.

Fig. 13 Field dependence of the magnetization at 5 K of $(La_{1-x}Pr_x)_{0.7}Ca_{0.3}MnO_3$ for a) $x=0.2$, b) $x=0.8$, c) $x=0.9$, and d) $x=1$. For large values of $x$, steps begin to appear in the magnetization.

Fig. 14 Doping dependence of the magnetic and resistive transitions for a) $(La_{1-x}Pr_x)_{0.7}Sr_{0.3}MnO_3$, b) $(La_{1-x}Pr_x)_{0.7}Ca_{0.3}MnO_3$ and c) $La_{1-y}Ba_yMnO_3$. It is clear that there is no systematic relationship between these transitions.

Table I

Magnetic parameters for $(La_{1-x}Pr_x)_{0.7}Sr_{0.3}MnO_3$

| $x$ | Saturation magnetization (emu/g) | Spontaneous magnetization (emu/g) |
|---|---|---|
| 0.2 | 91 | 90 |
| 0.3 | 92 | 92 |
| 0.4 | 92 | 92 |
| 0.5 | 93 | 92 |
| 0.6 | 94 | 92 |
| 0.8 | 95 | 93 |
| 1.0 | 99 | 95 |

Table II
Magnetic parameters for $(La_{1-x}Pr_x)_{0.7}Ca_{0.3}MnO_3$

| $x$ | Saturation Magnetization (emu/g) | Spontaneous Magnetization (emu/g) |
|---|---|---|
| 0.2 | 95 | 95 |
| 0.4 | 97 | 94 |
| 0.5 | 98 | 101 |
| 0.6 | 104 | 100 |
| 0.7 | 105 | 101 |
| 0.8 | 109 | 102 |
| 0.9 | 110 | 102 |
| 1.0 | 120 | - |

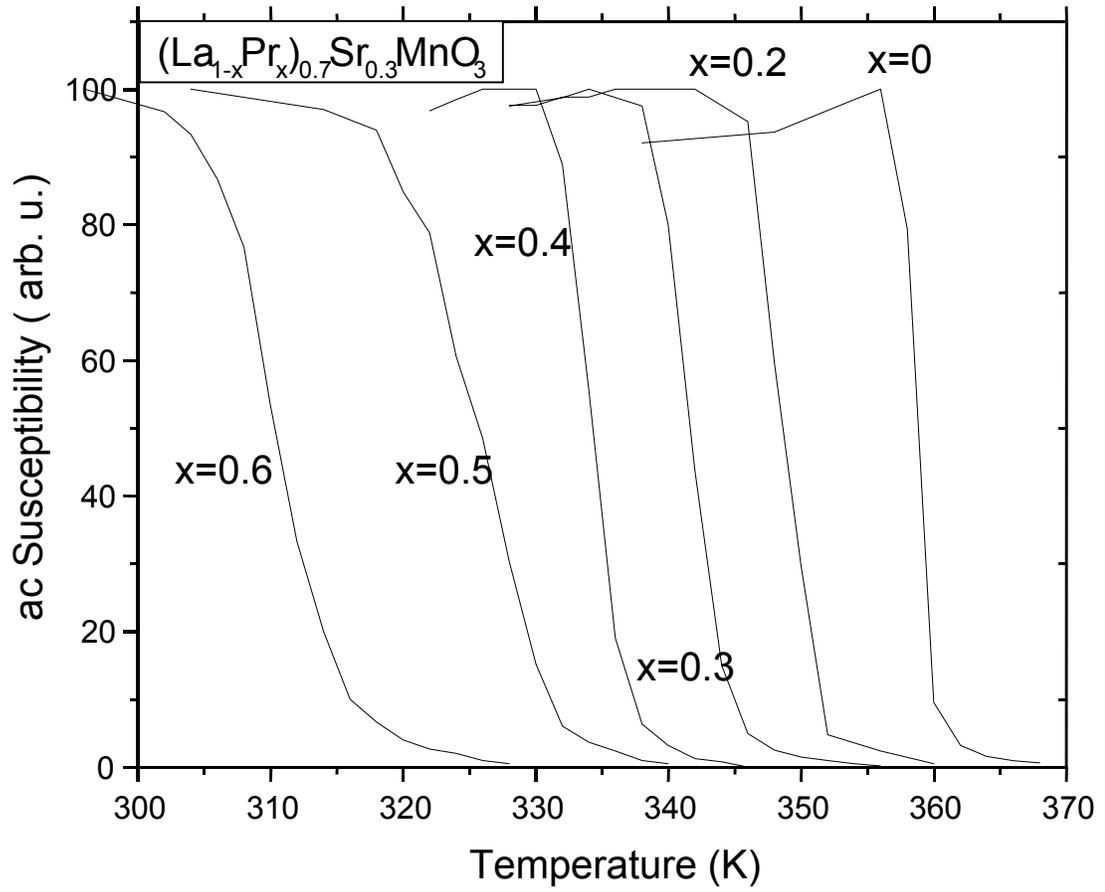

Figure 1a

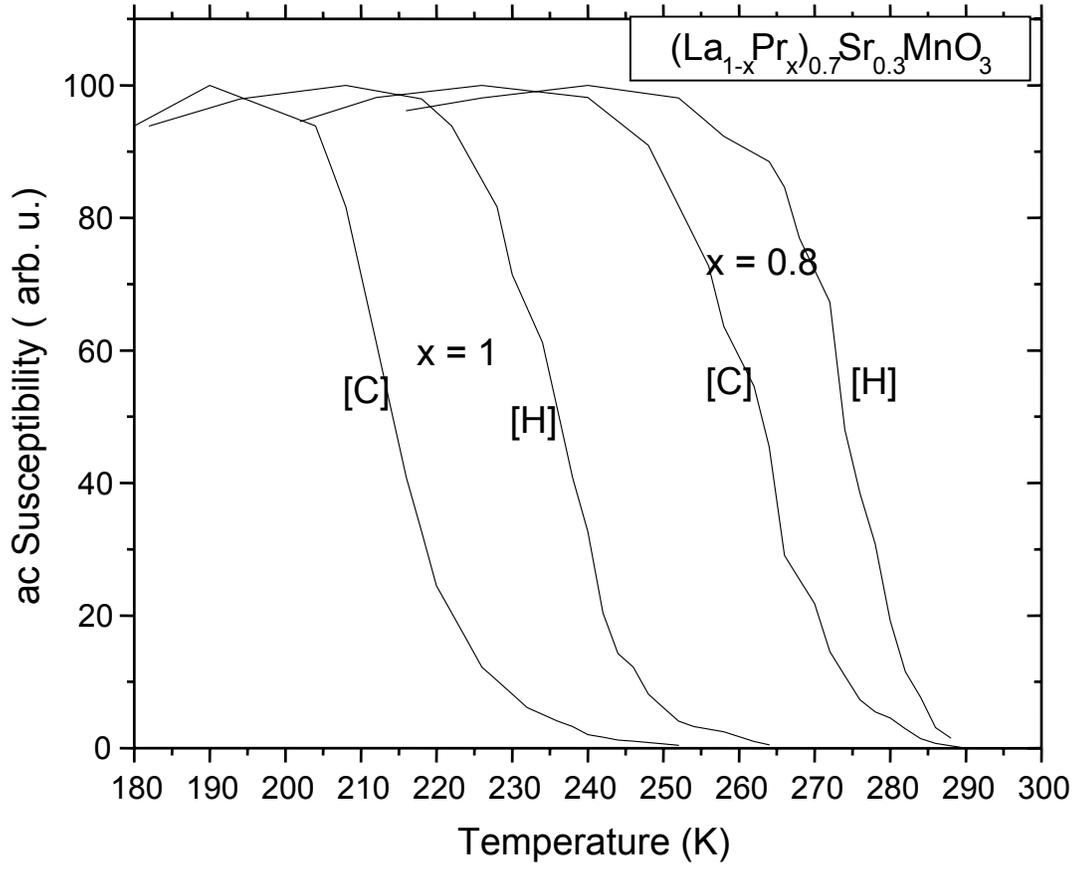

Figure 1b

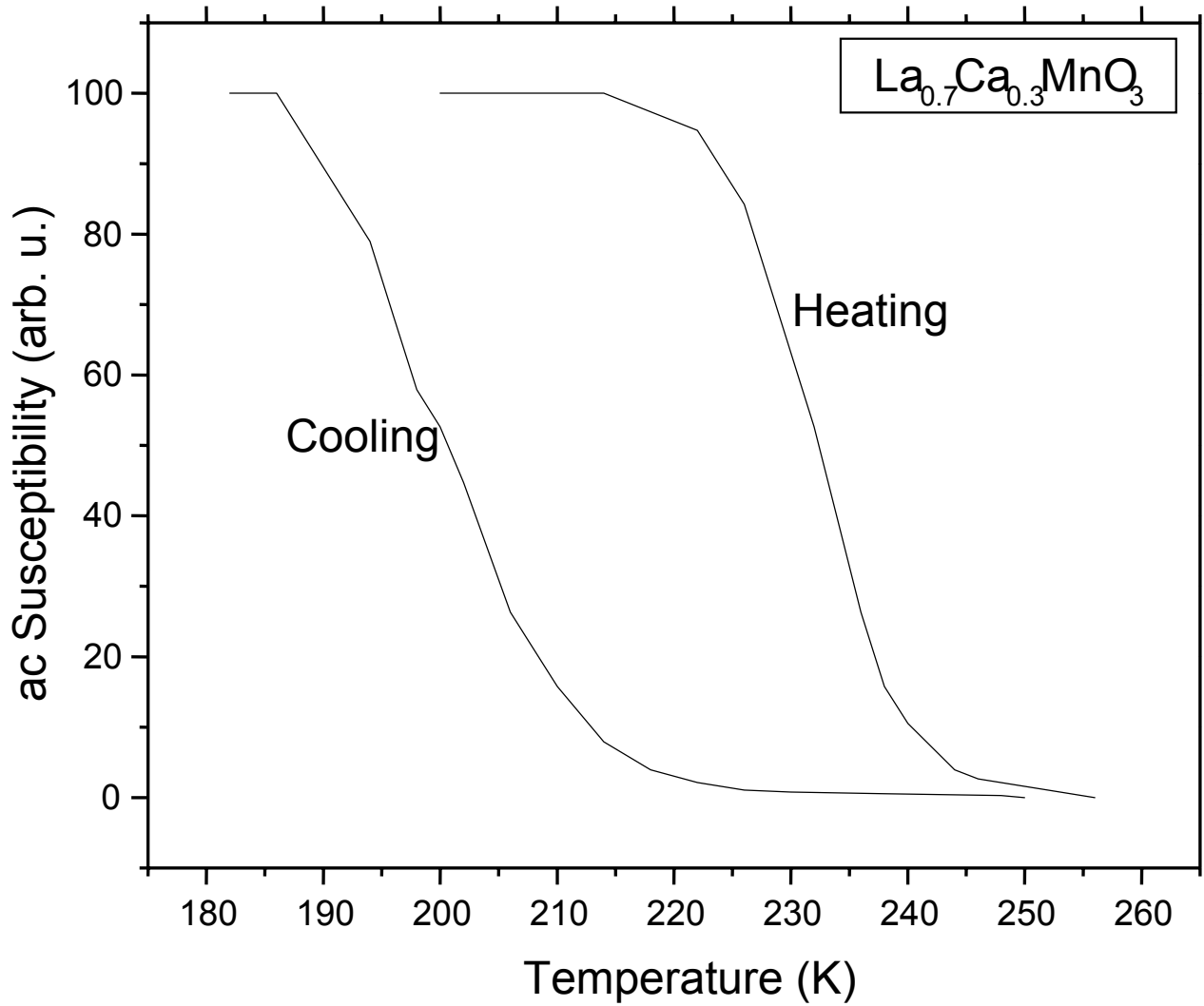

Figure 2a

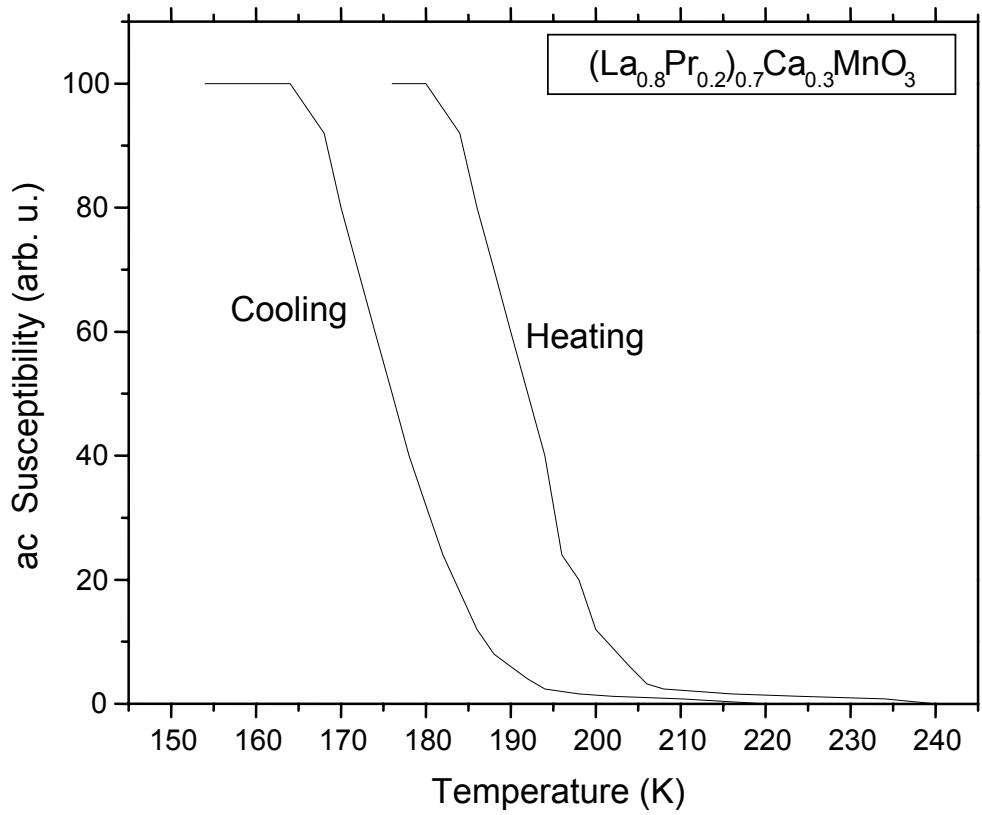

Figure 2b

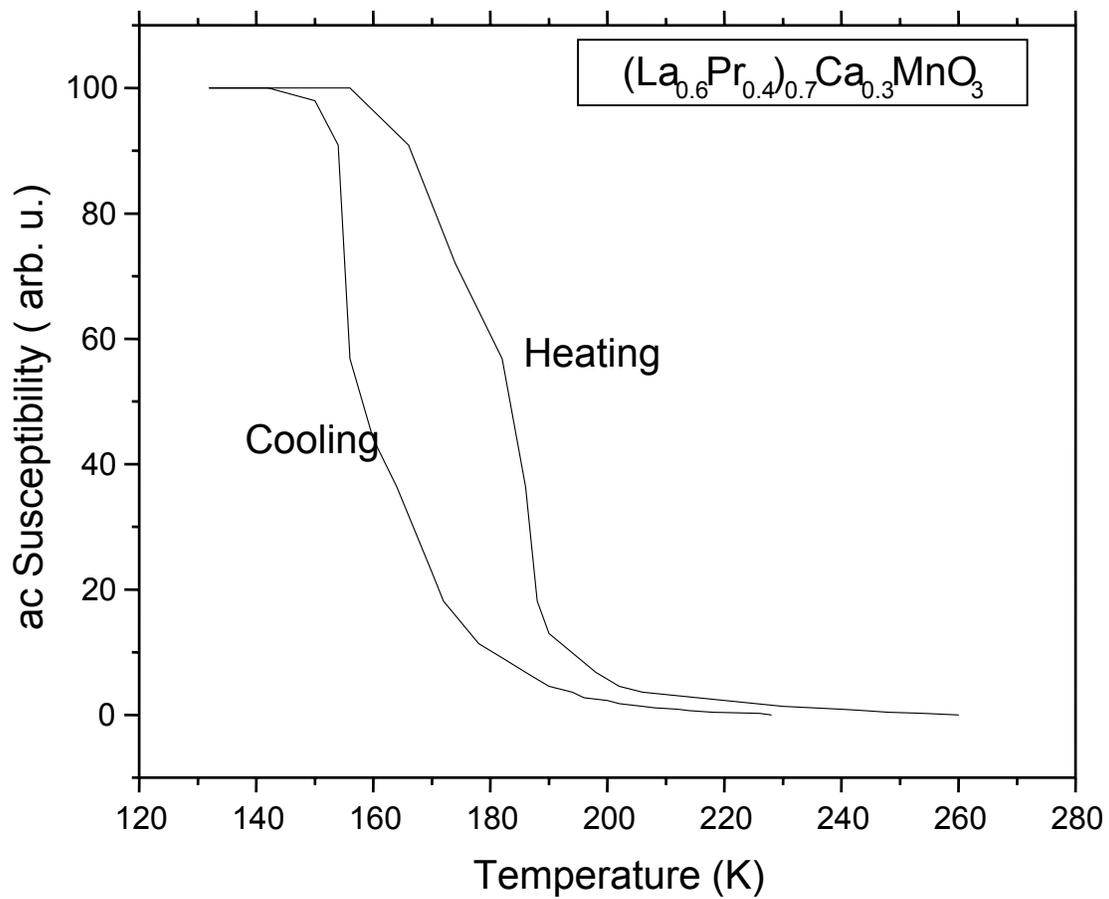

Figure 2c

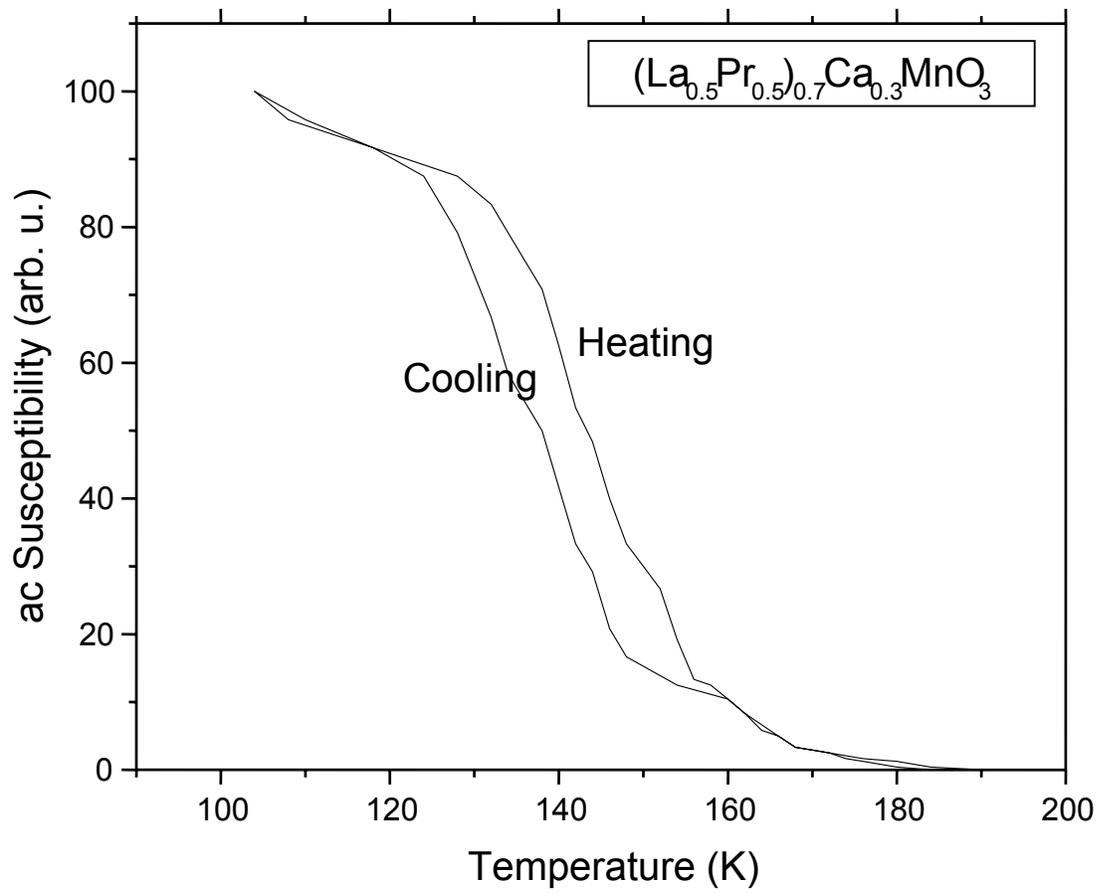

Figure 2d

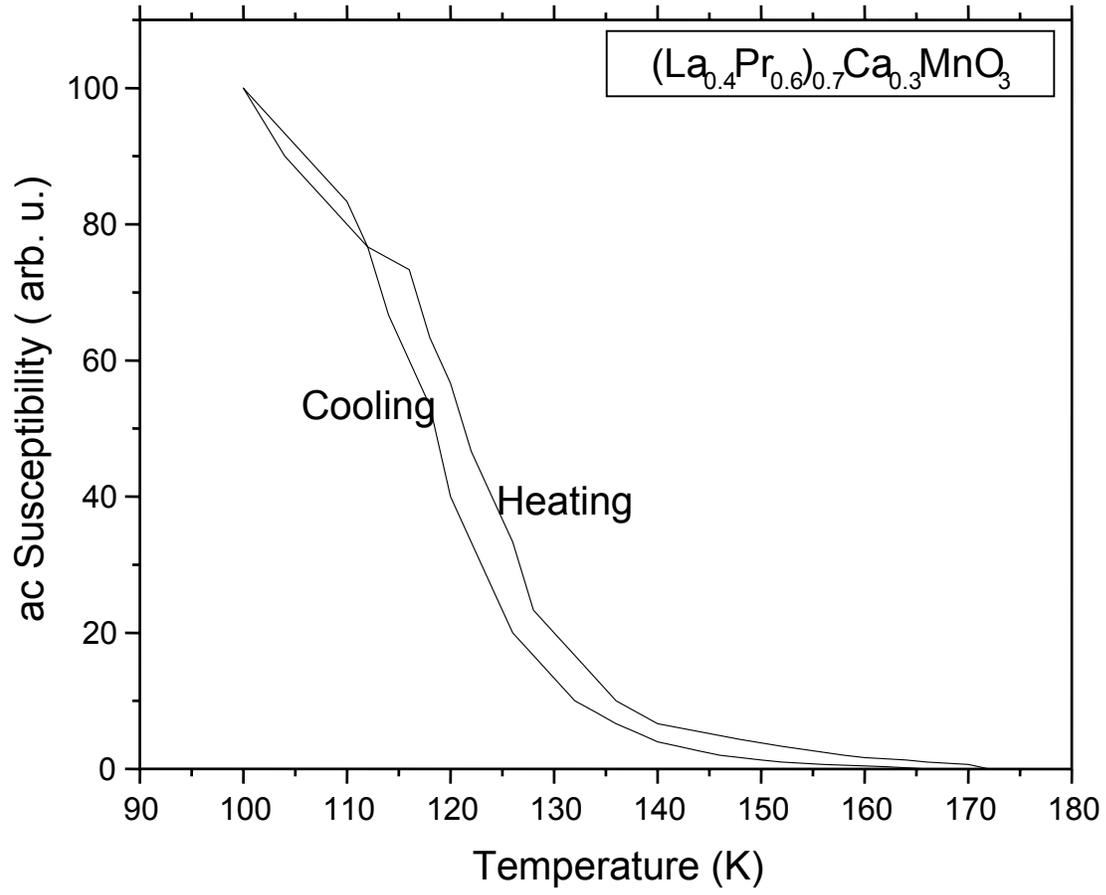

Figure 2e

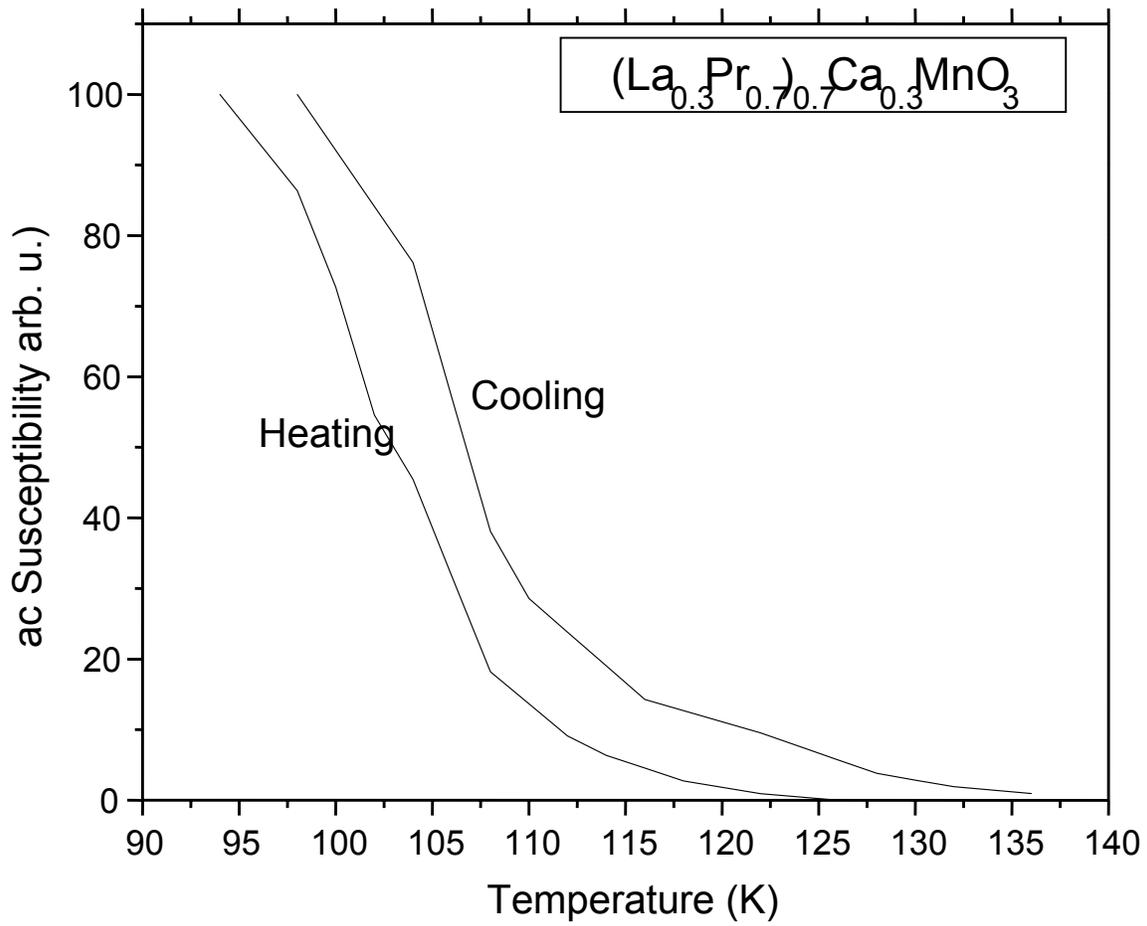

Figure 2f

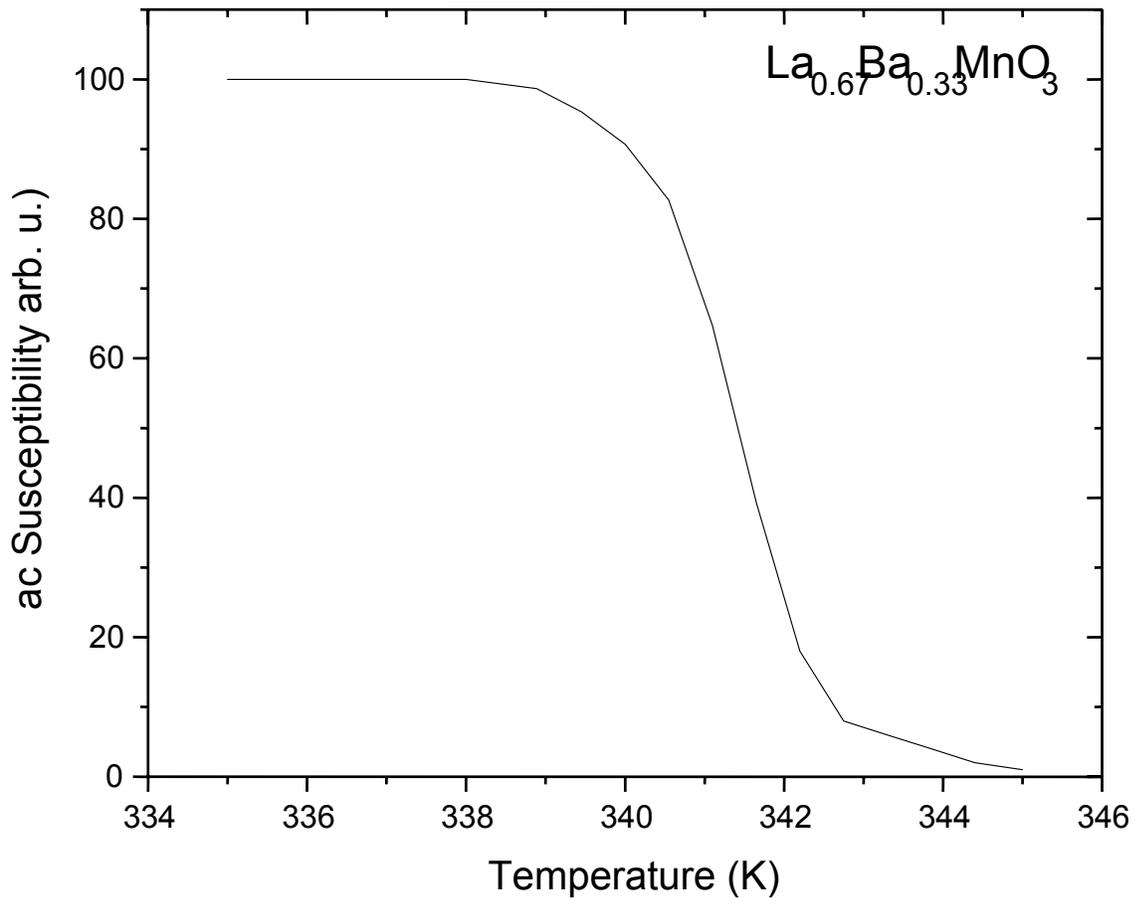

Figure 3a

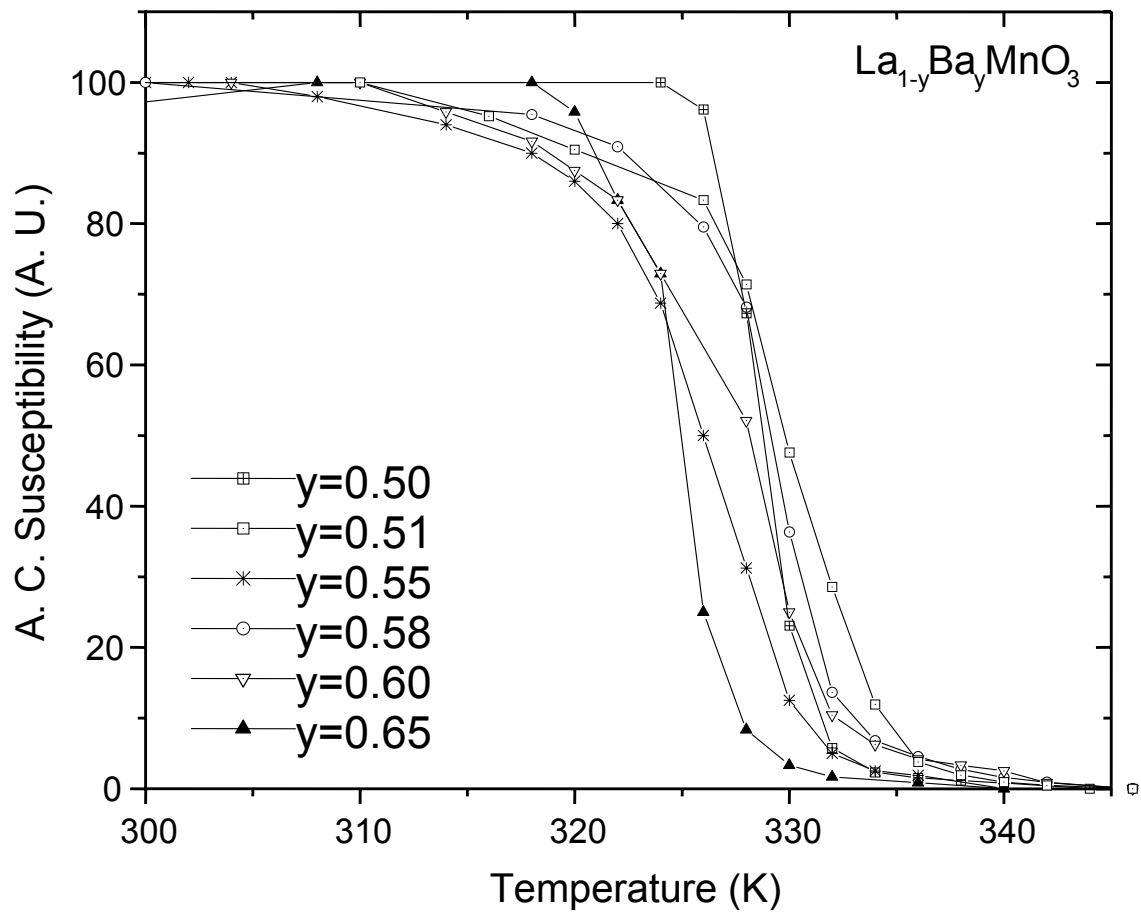

Figure 3b

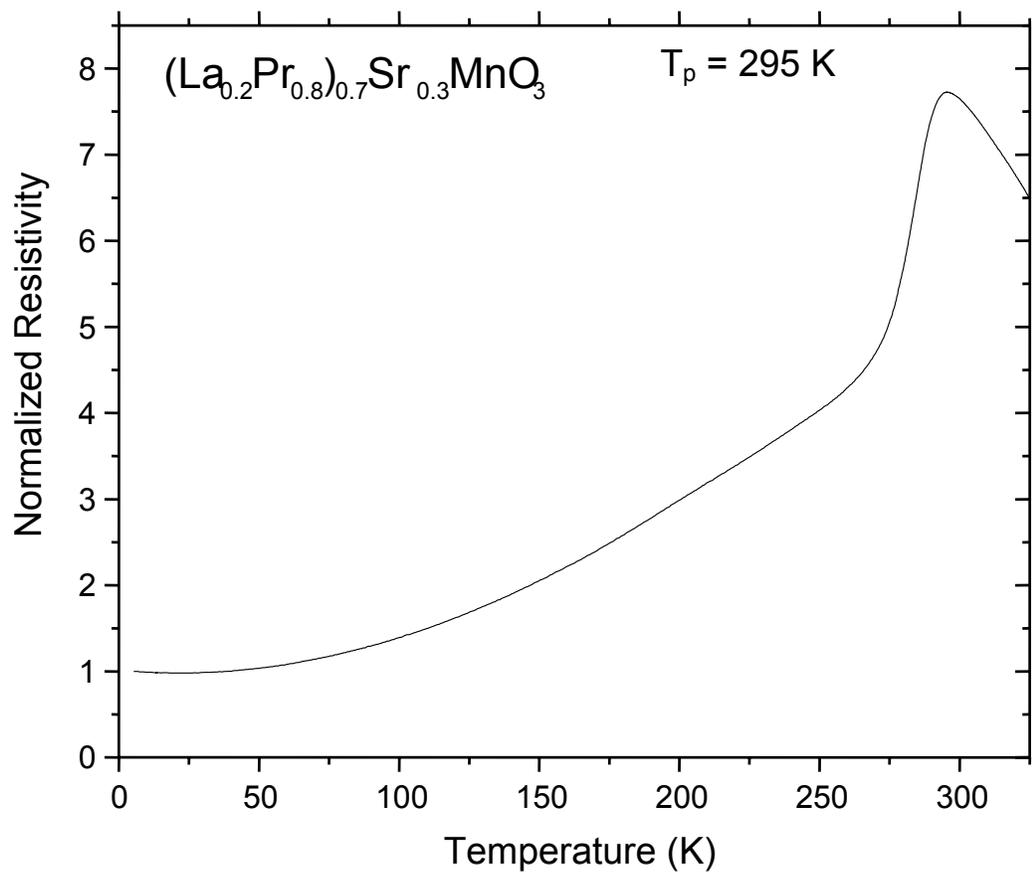

Figure 4a

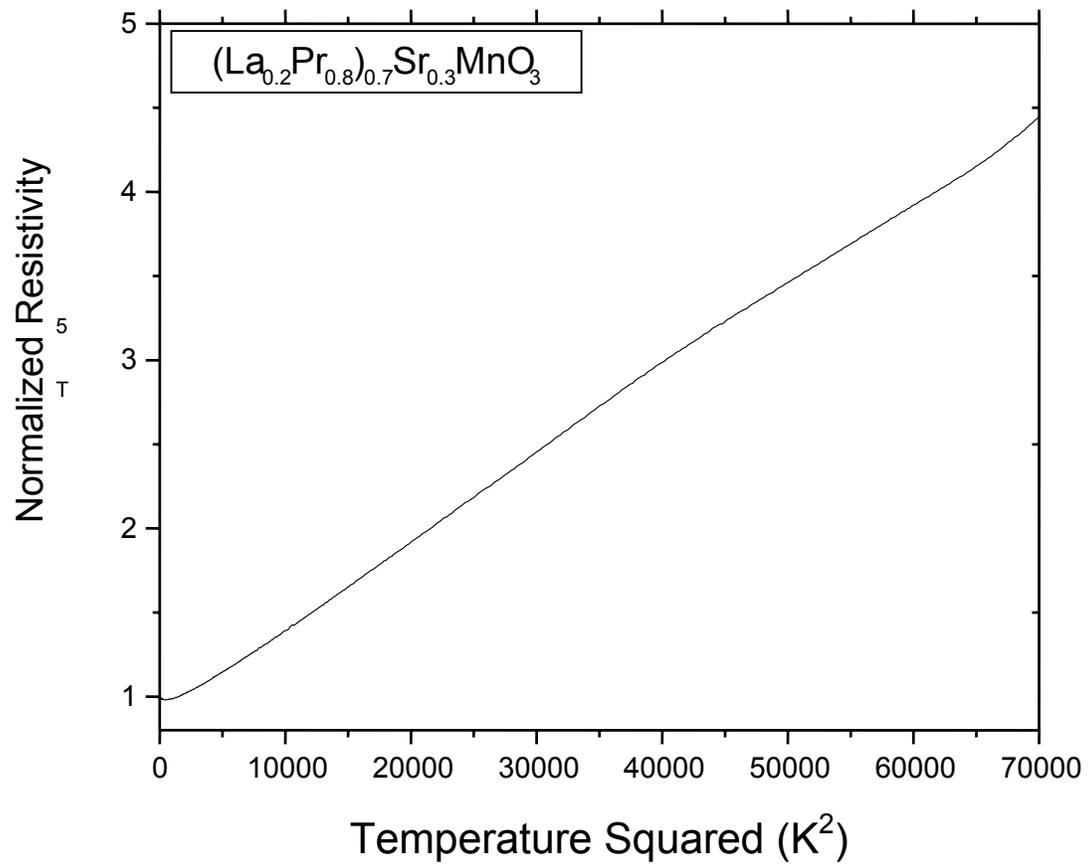

Figure 4b

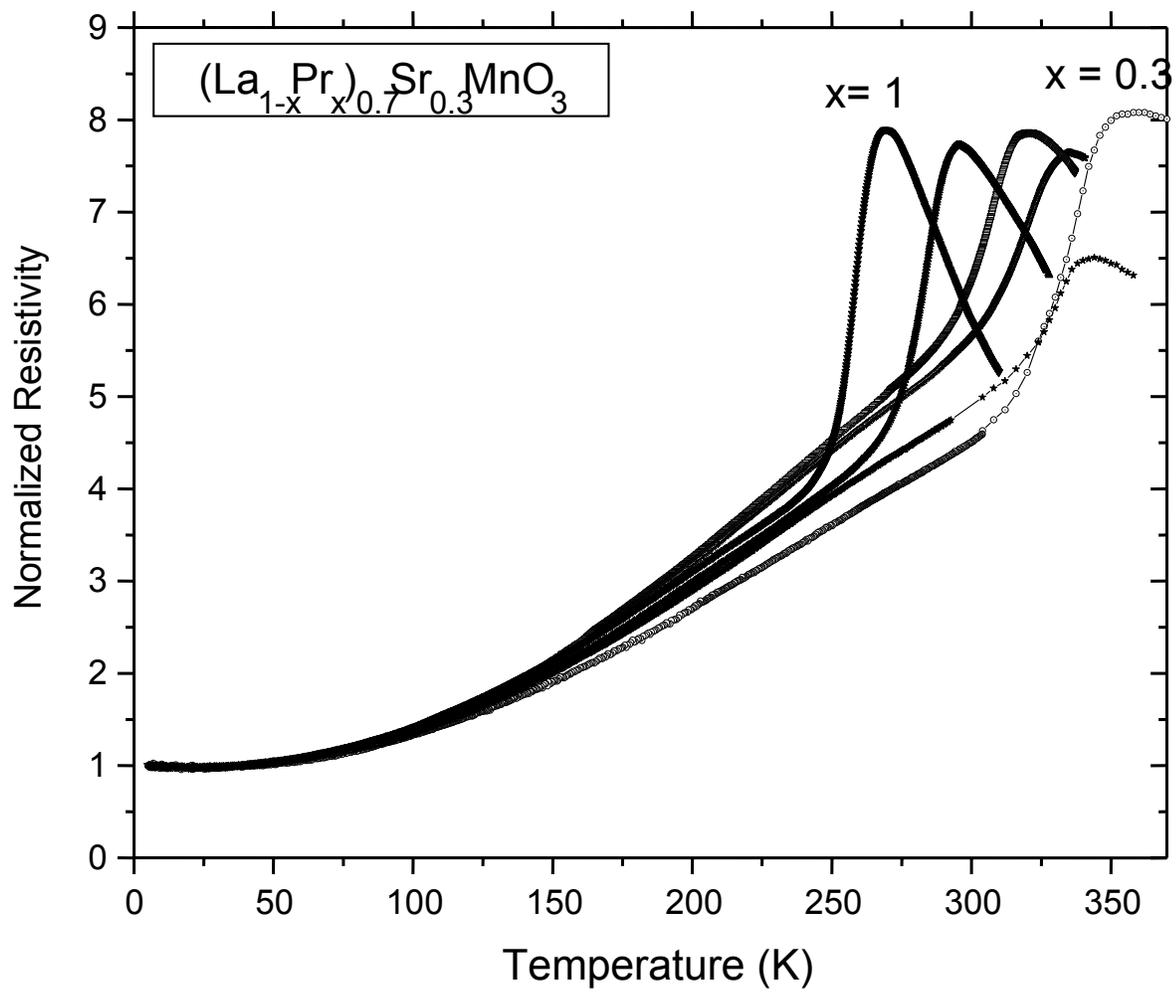

Figure 4c

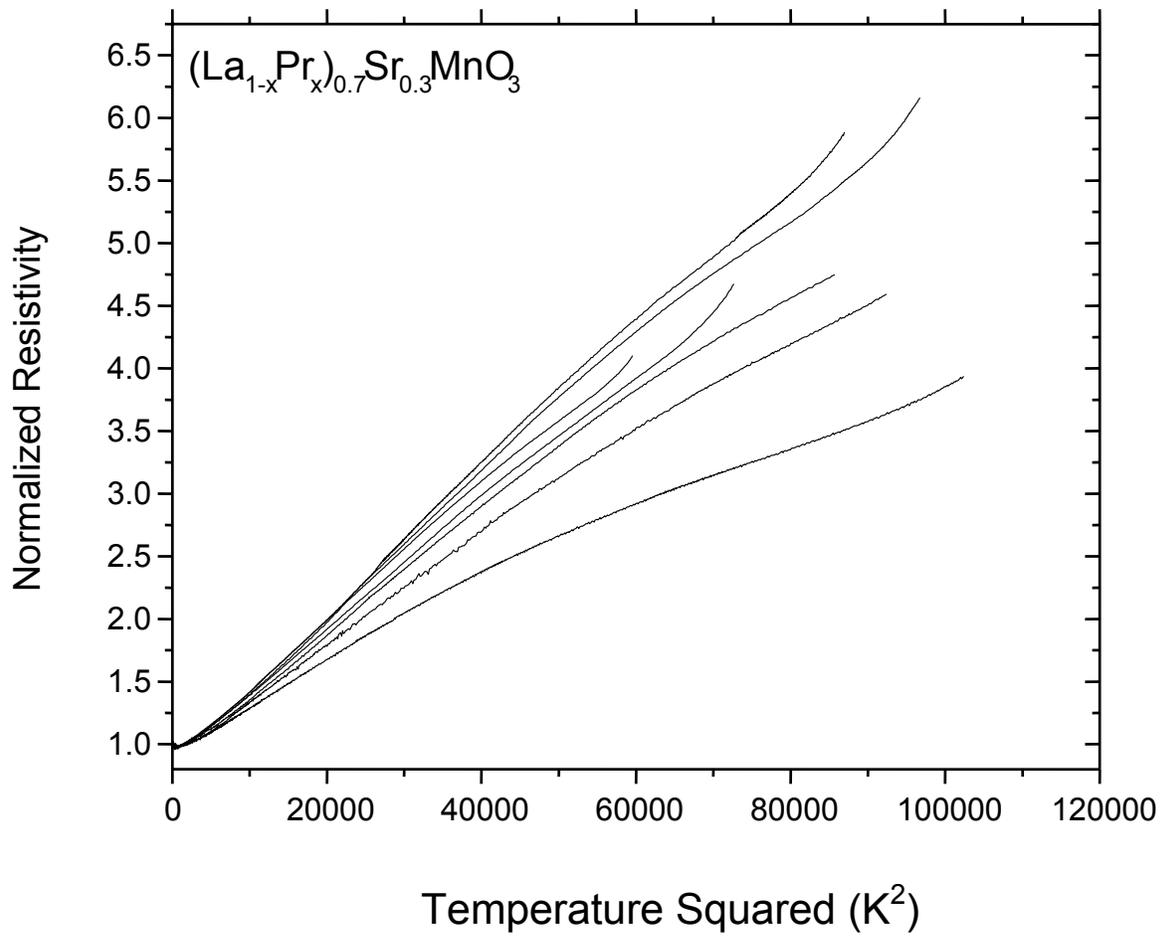

Figure 4d

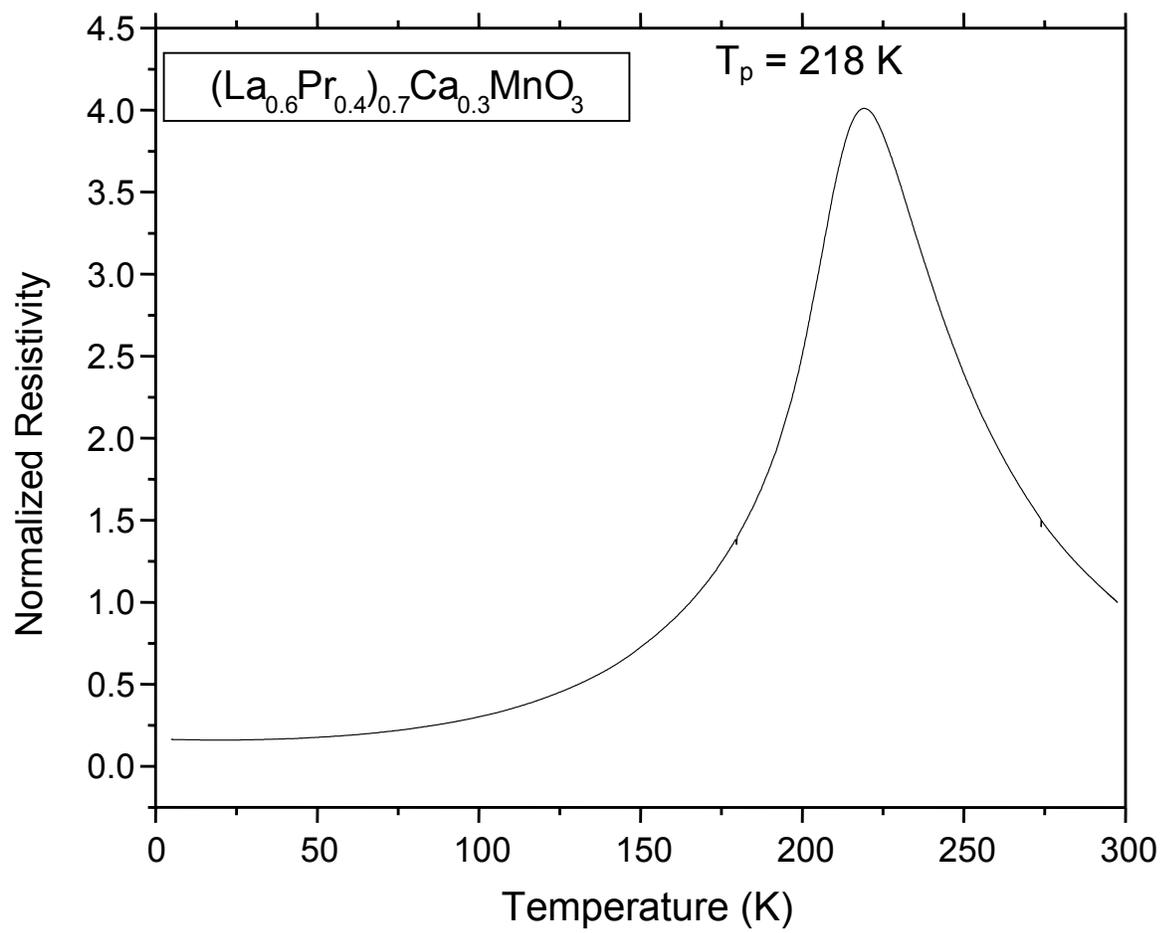

Figure 5a

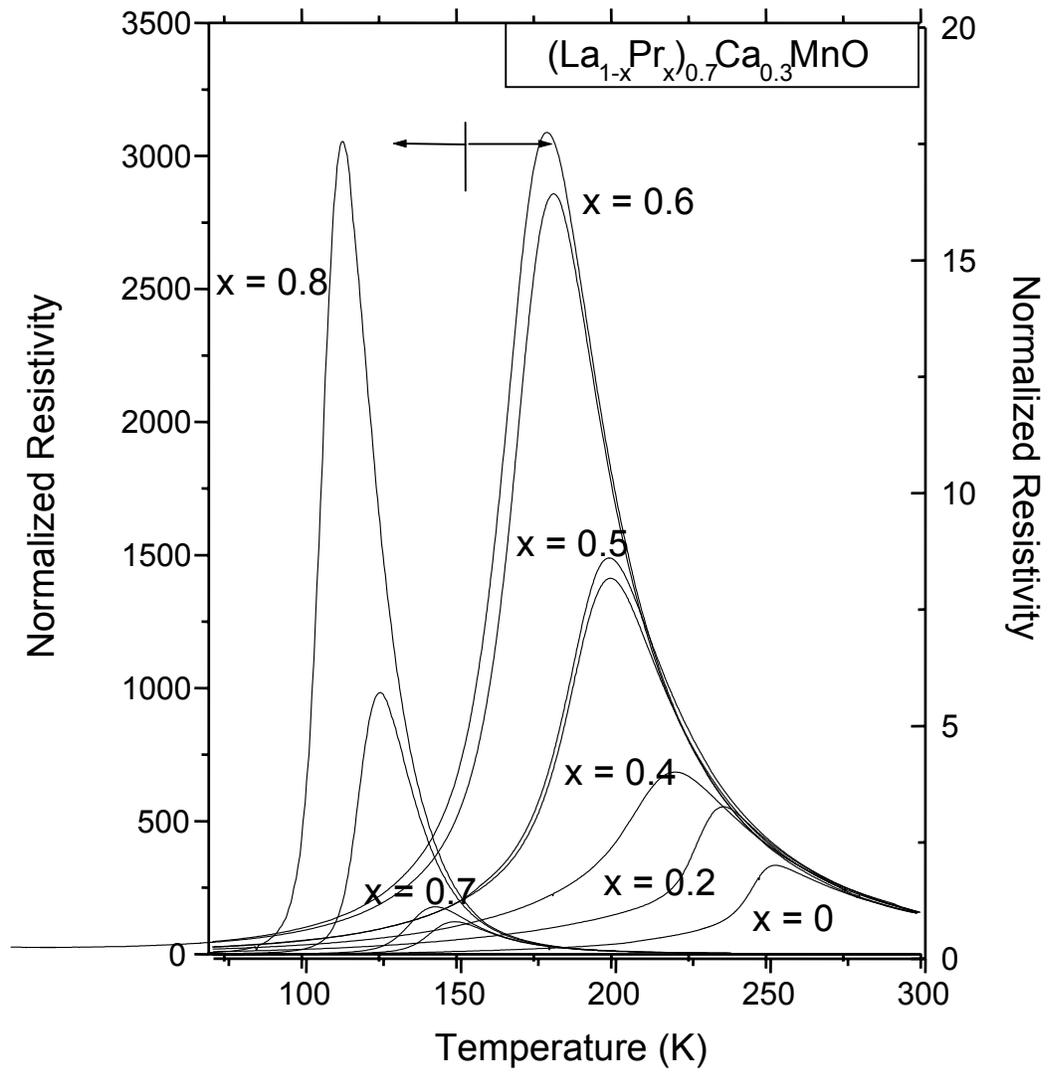

Figure 5b

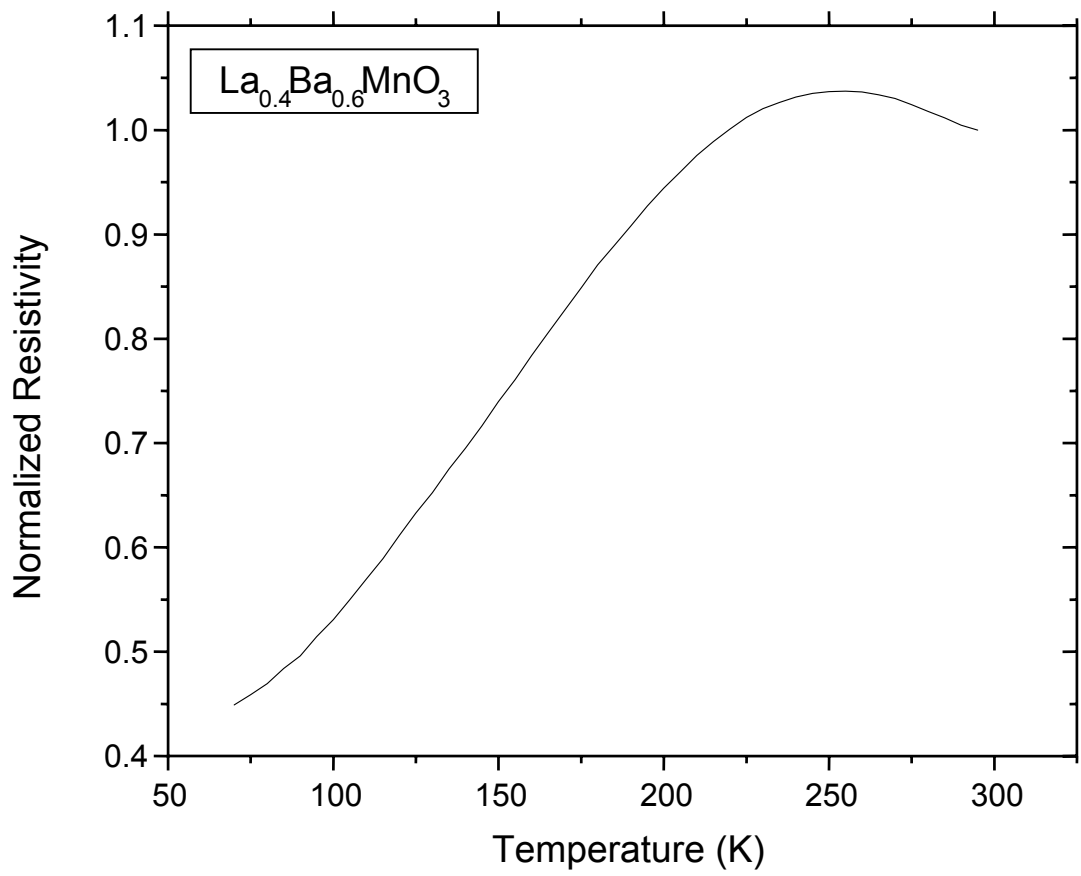

Figure 6

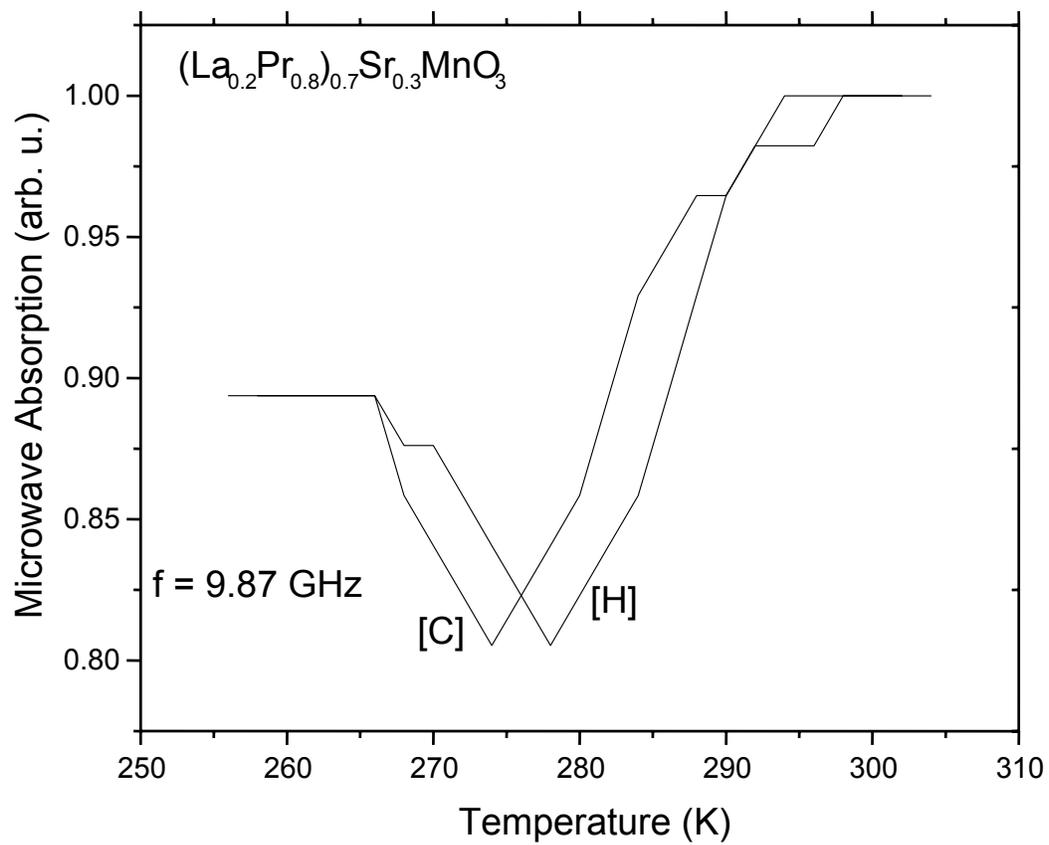

Figure 7a

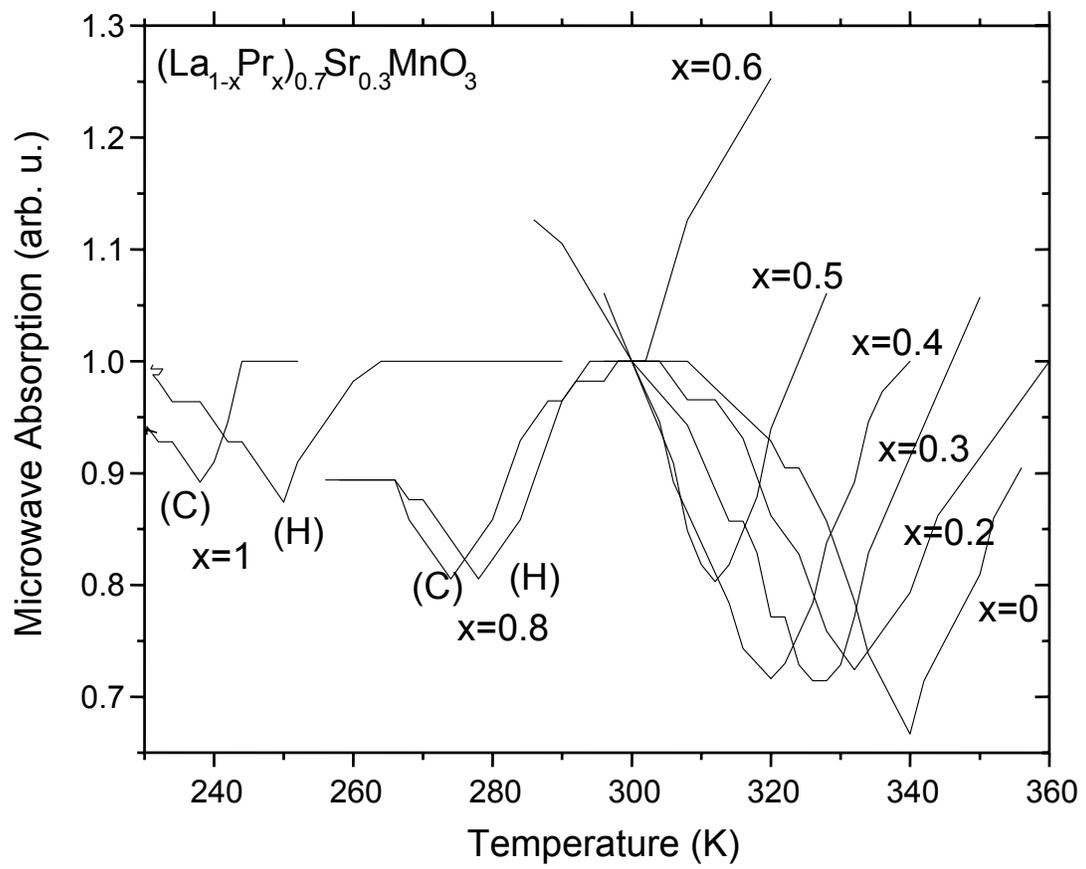

Figure 7b

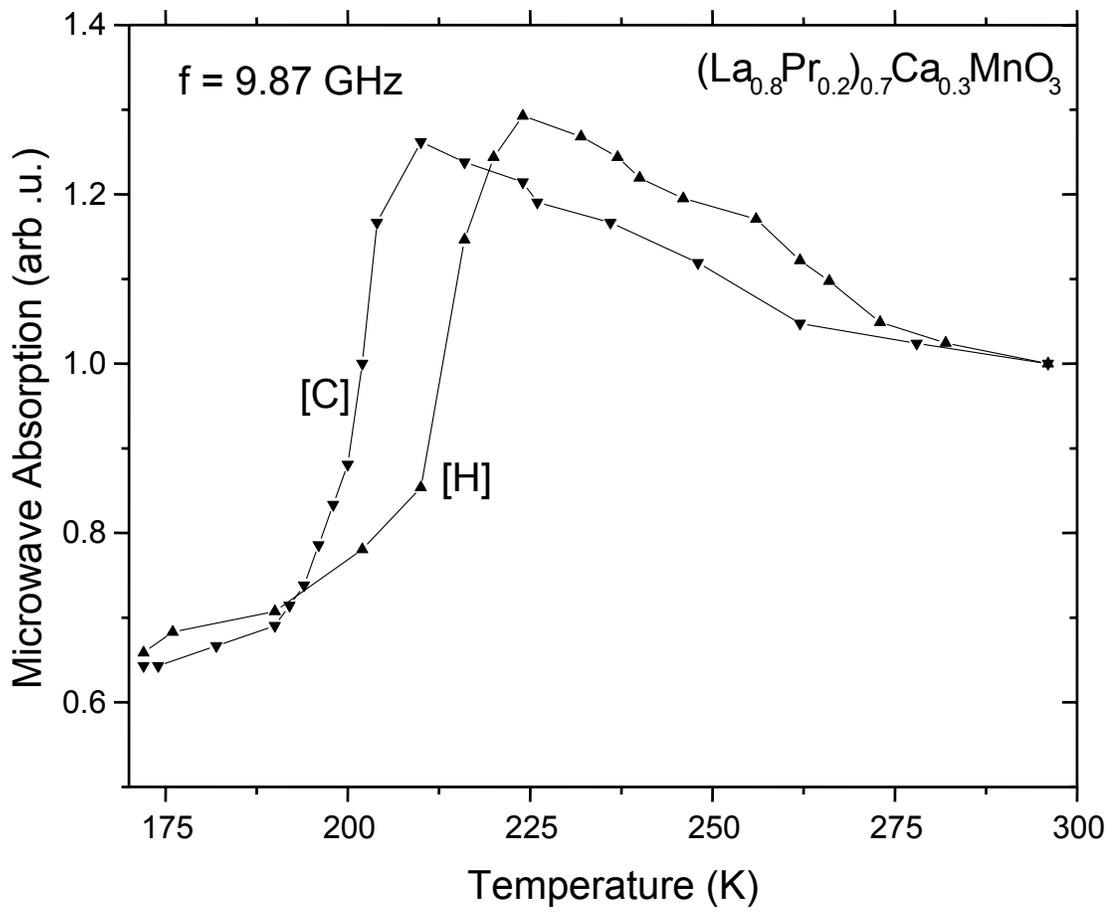

Figure 8a

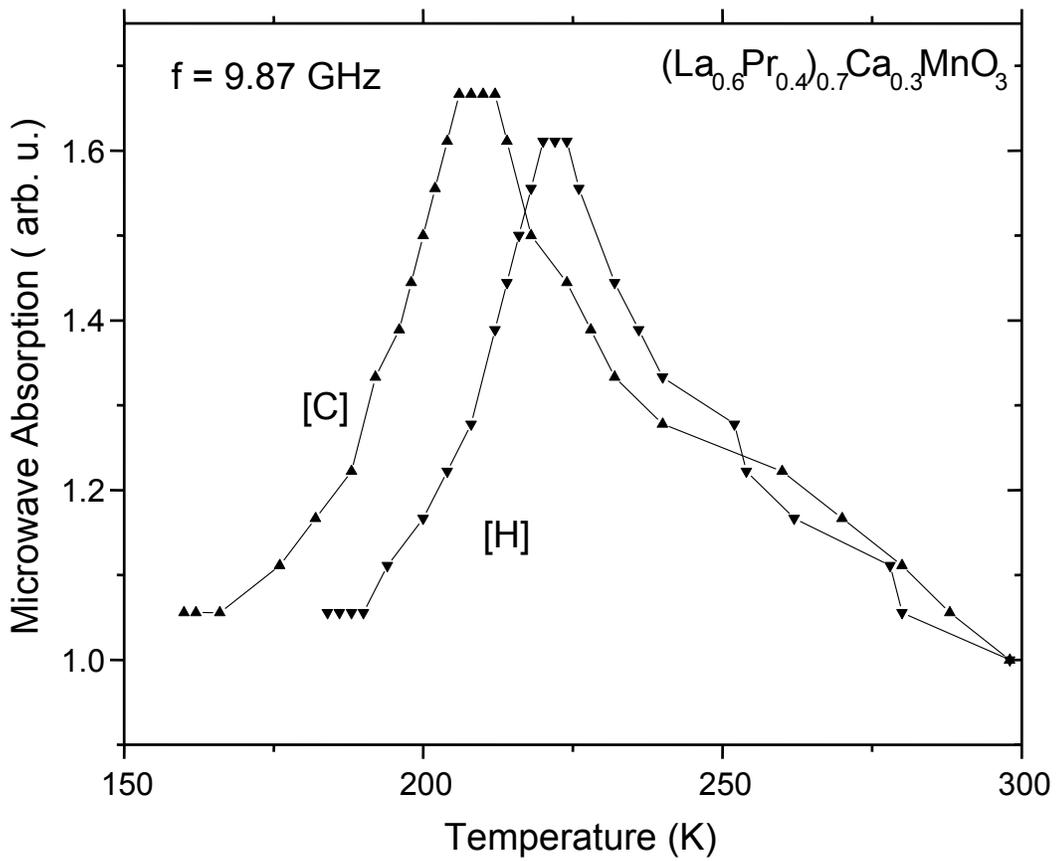

Figure 8b

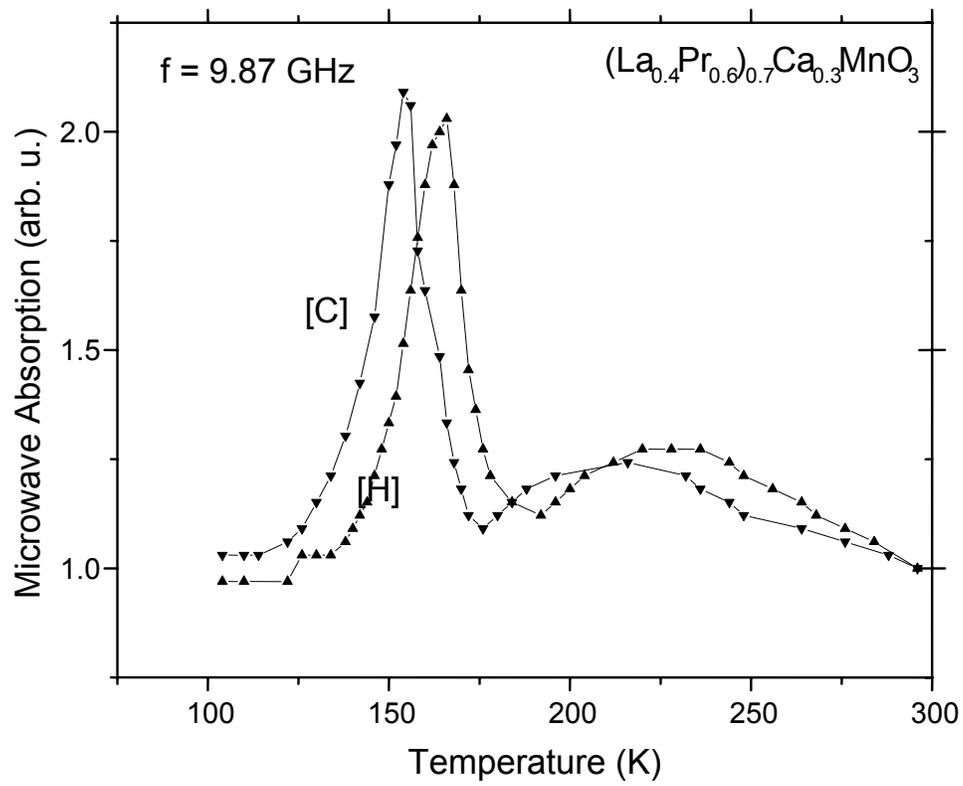

Figure 9a

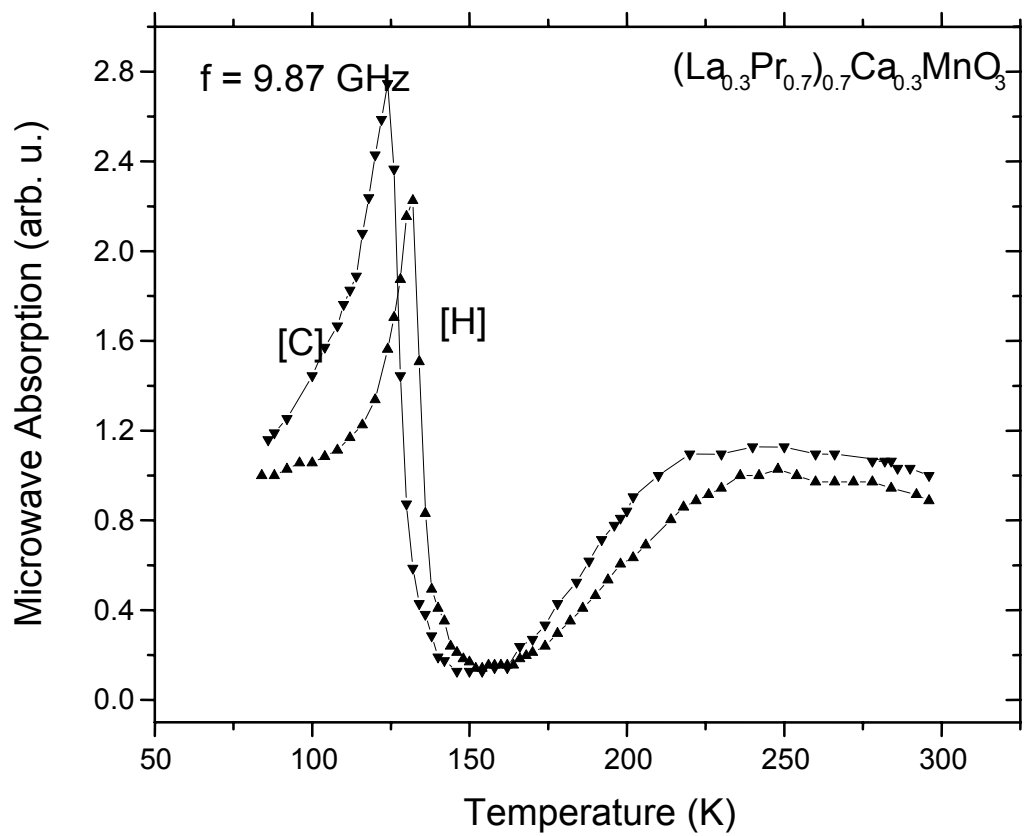

Figure 9b

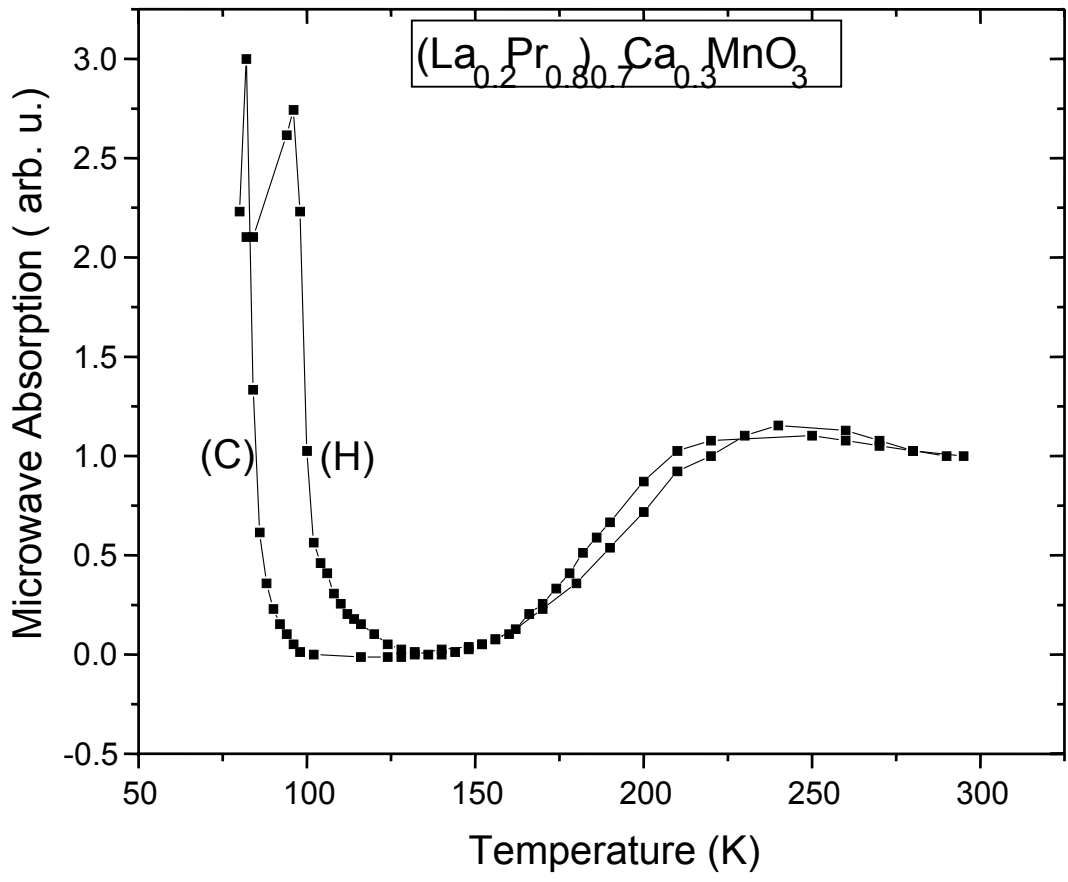

Figure 9c

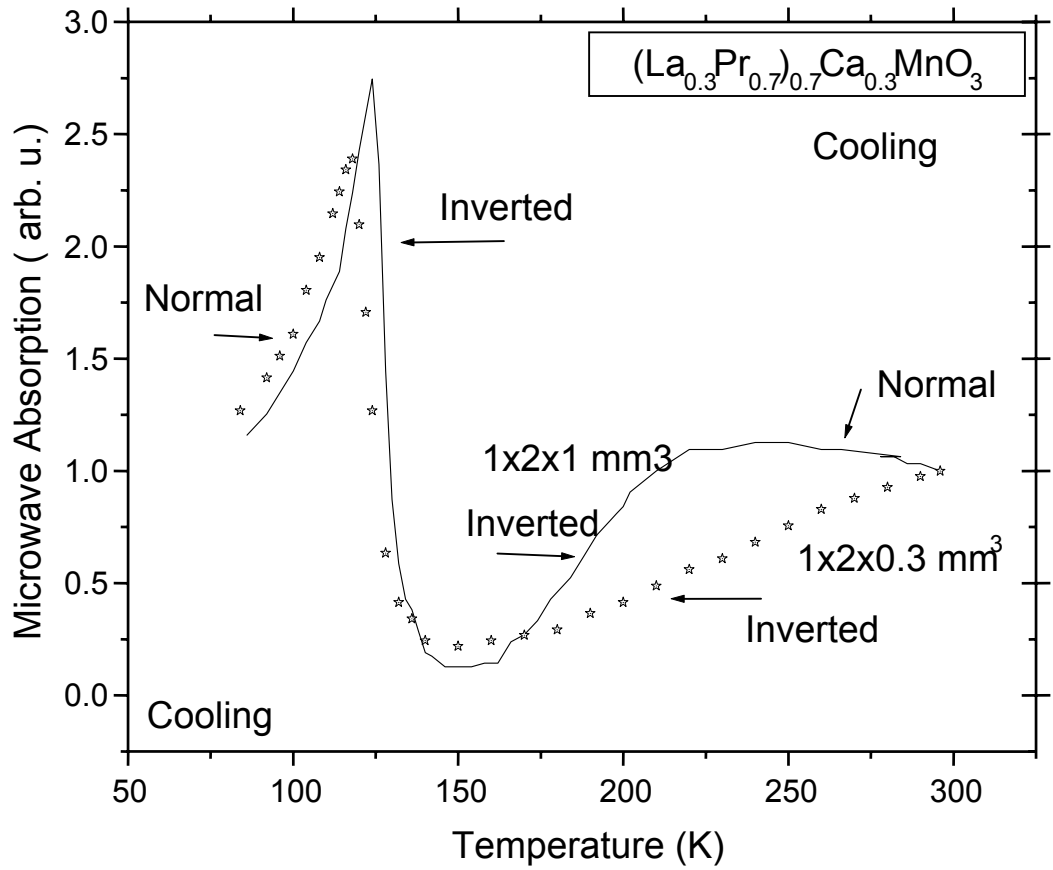

Figure 10

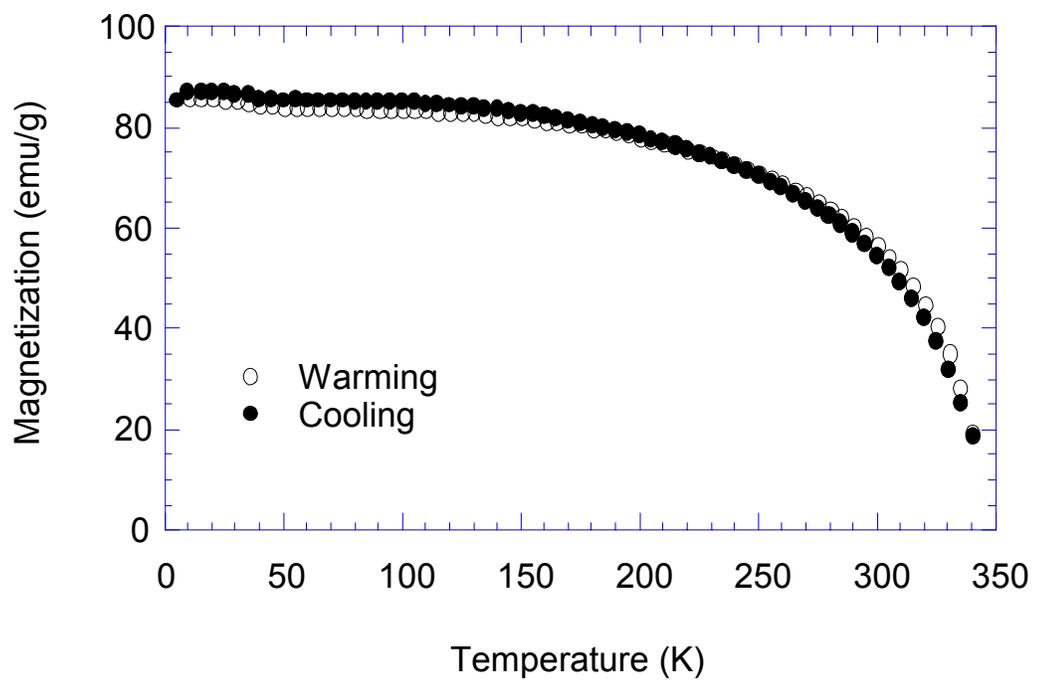

Figure 11a

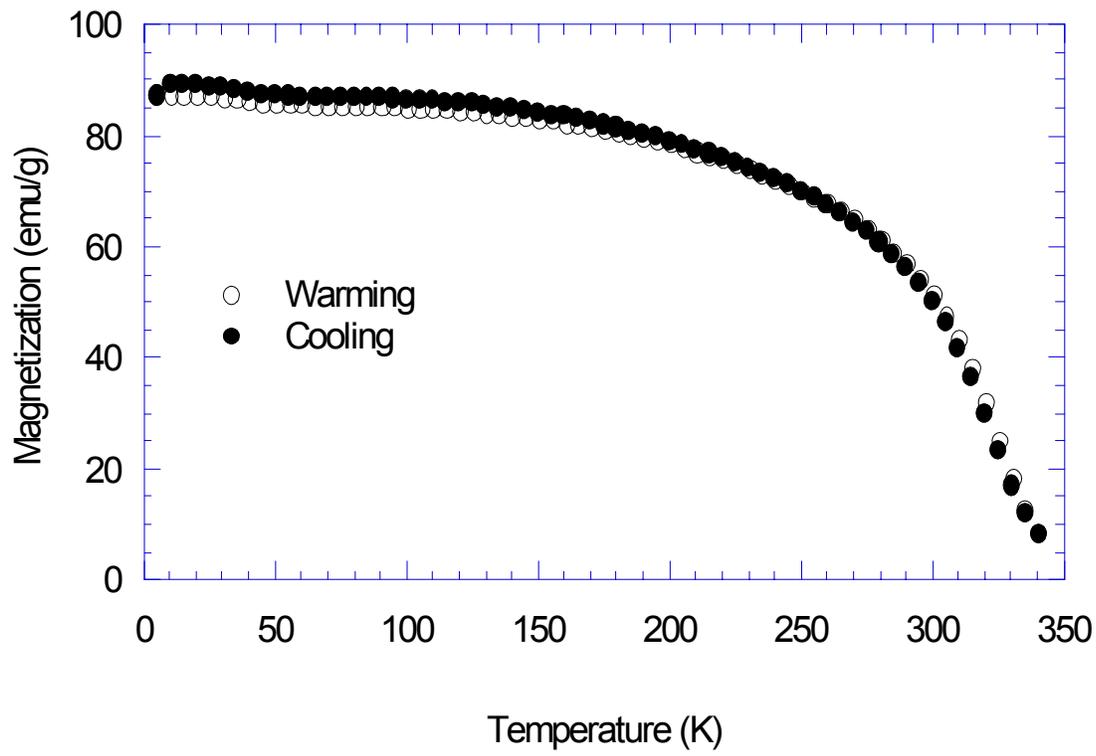

Figure11b

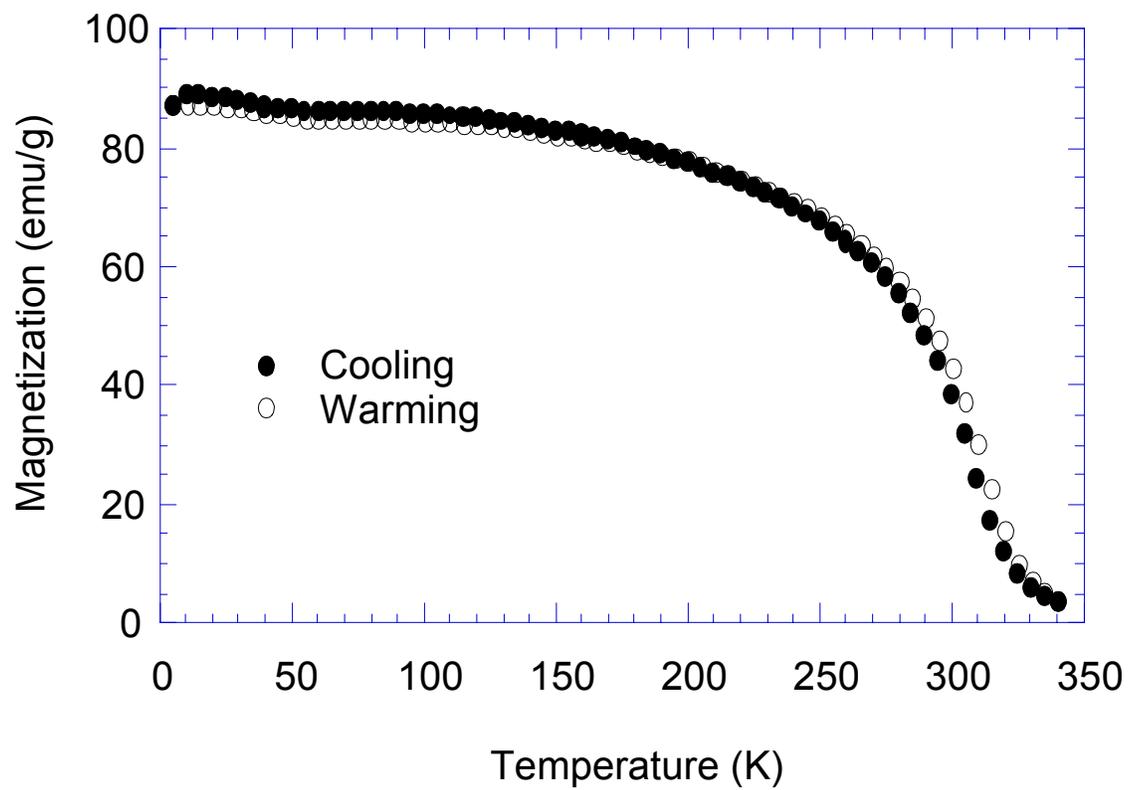

Figure 11c

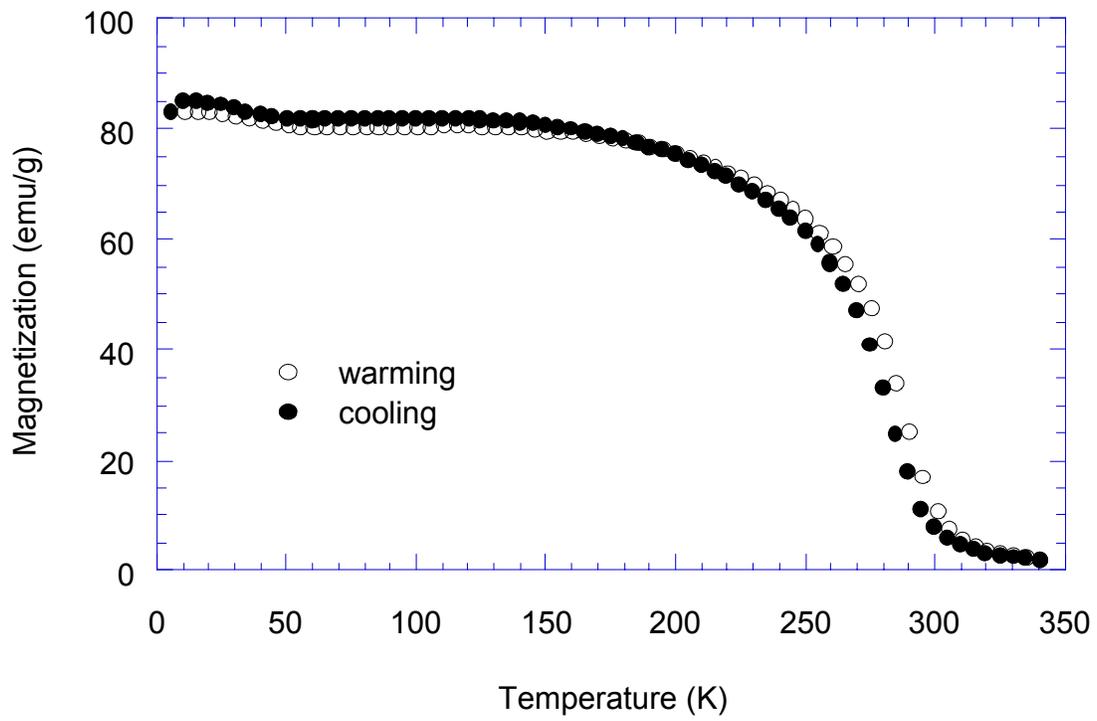

Figure 11d

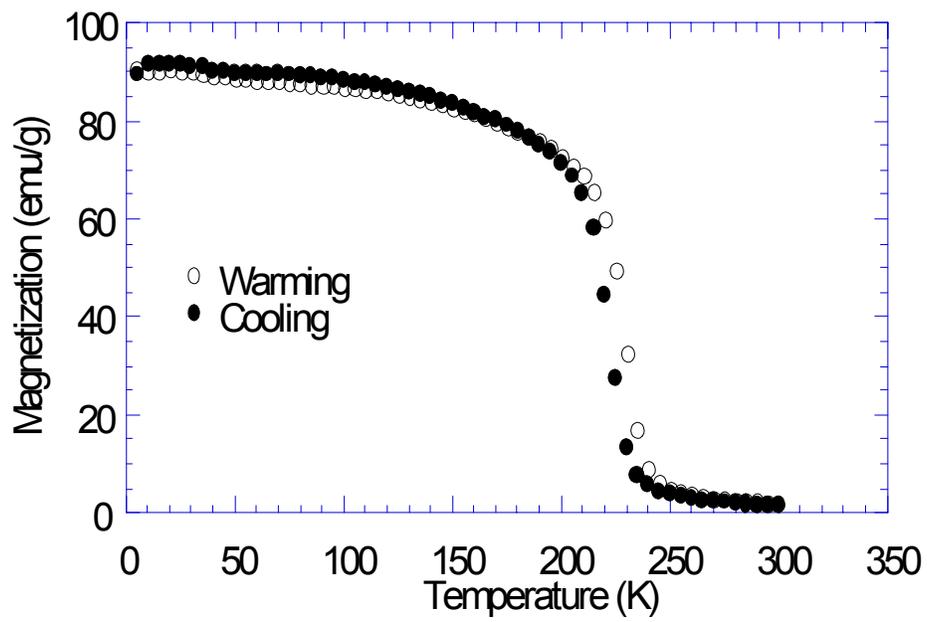

Figure 12a

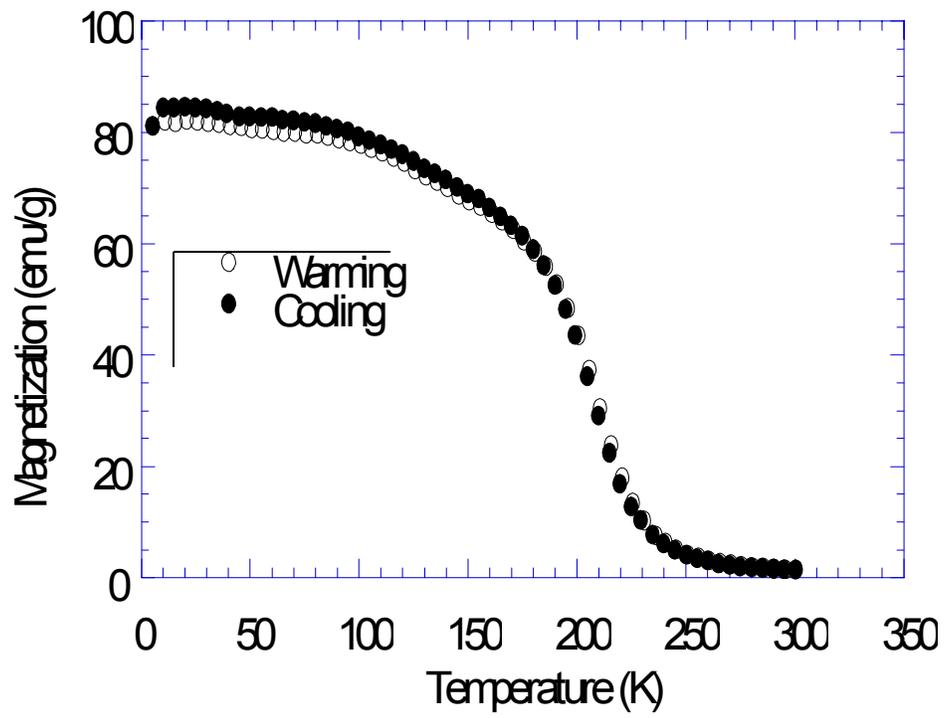

Figure 12b

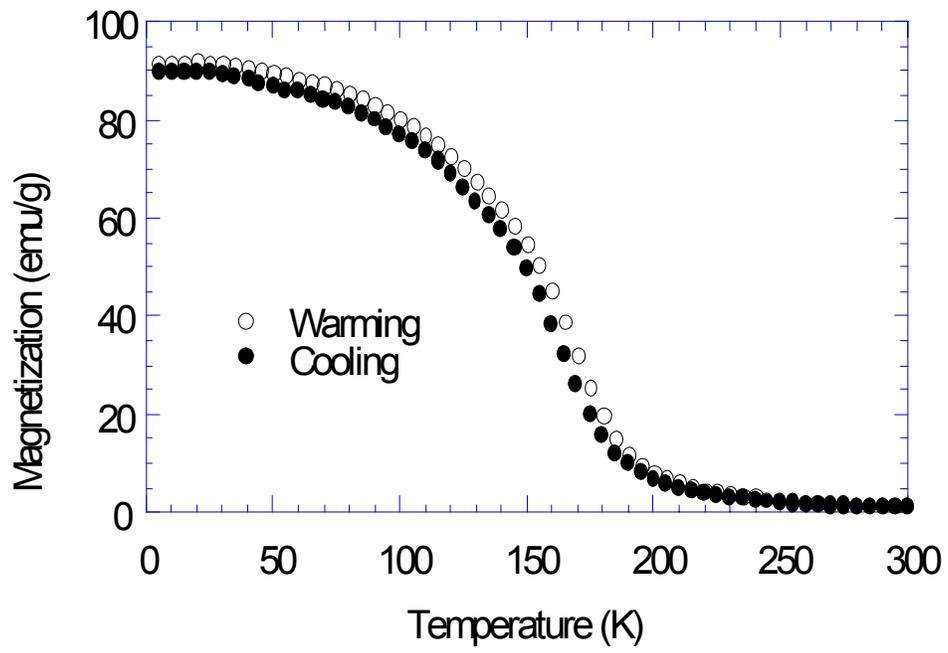

Figure 12c

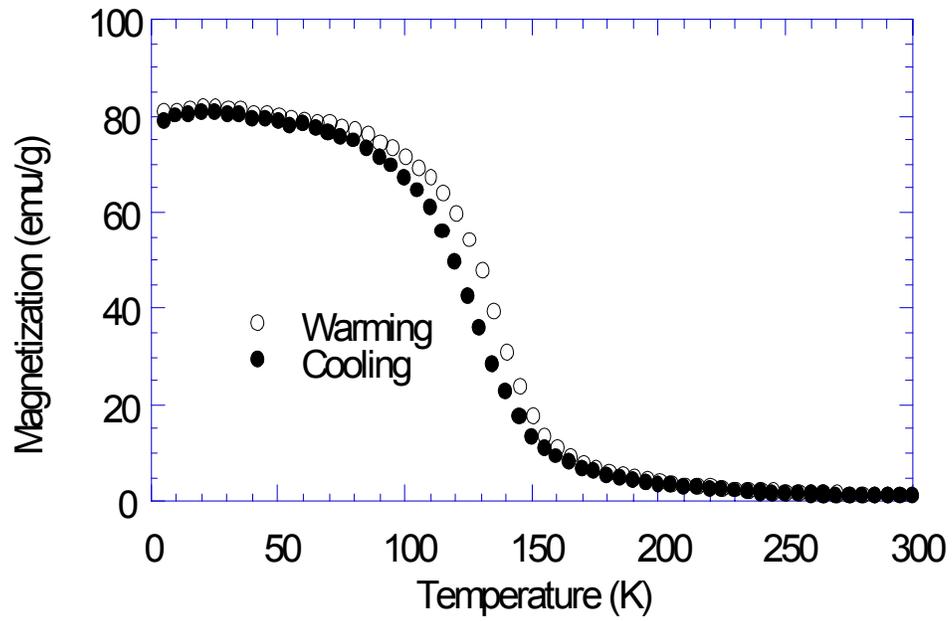

Figure 12d

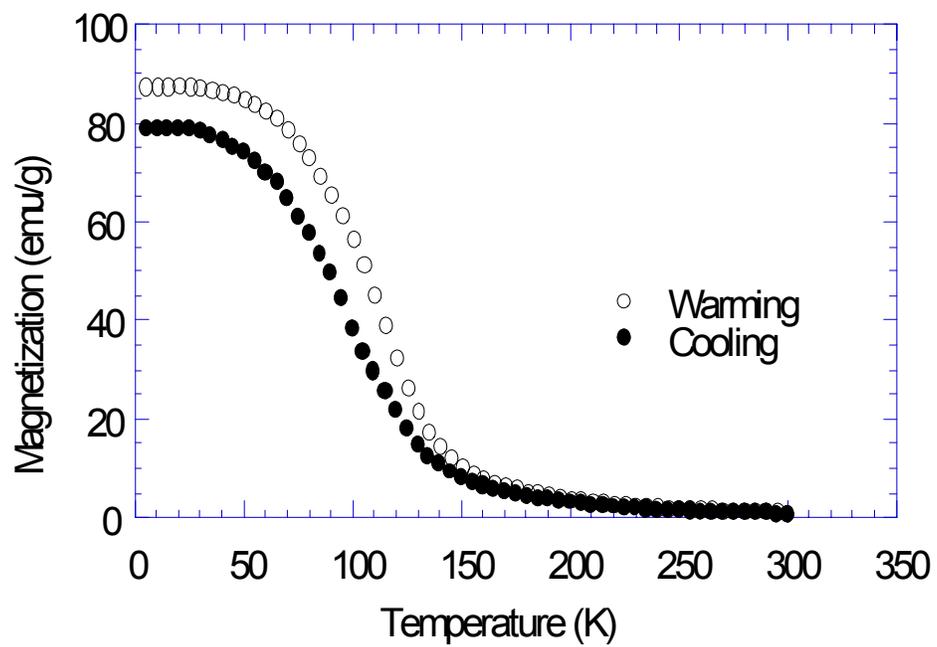

Figure 12e

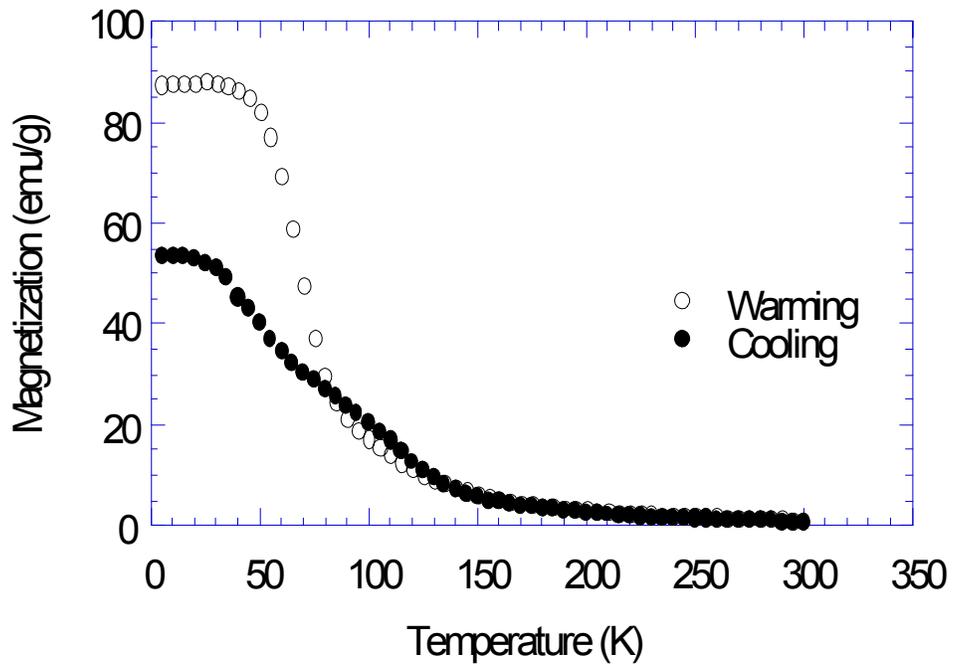

Figure 12f

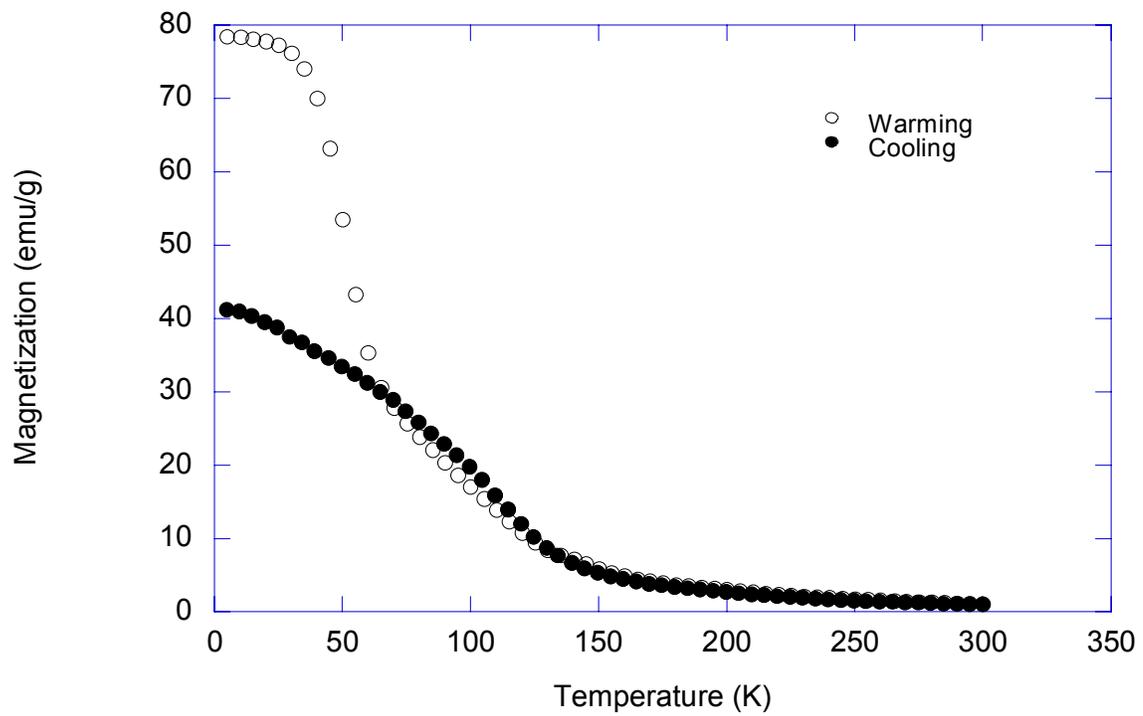

Figure 12g

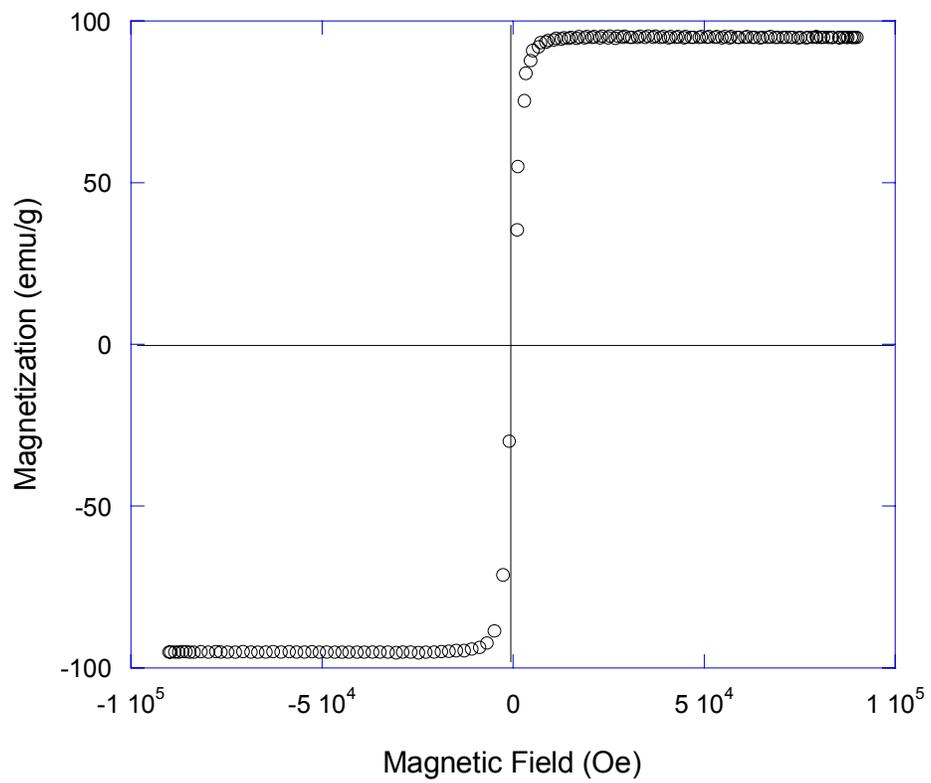

Figure 13a

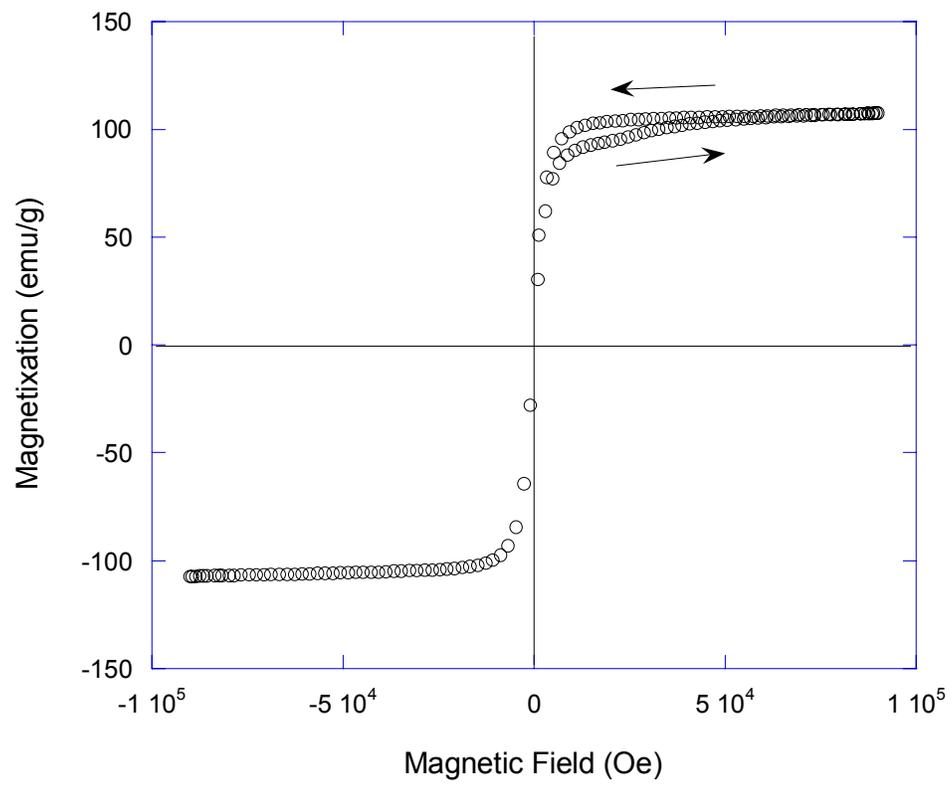

Figure 13b

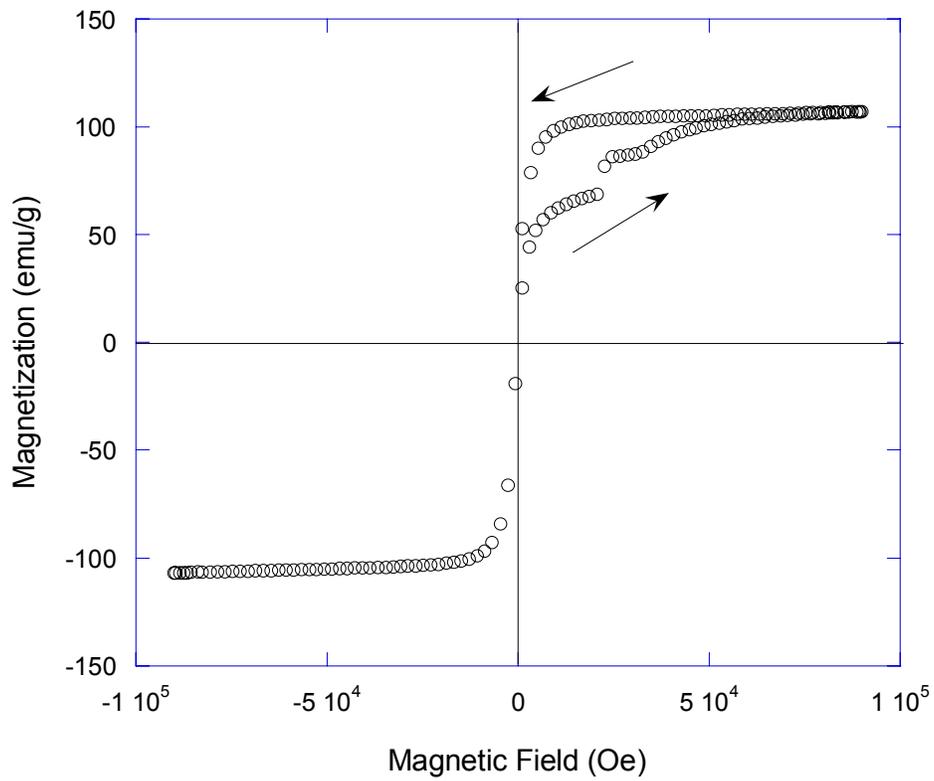

Figure 13c

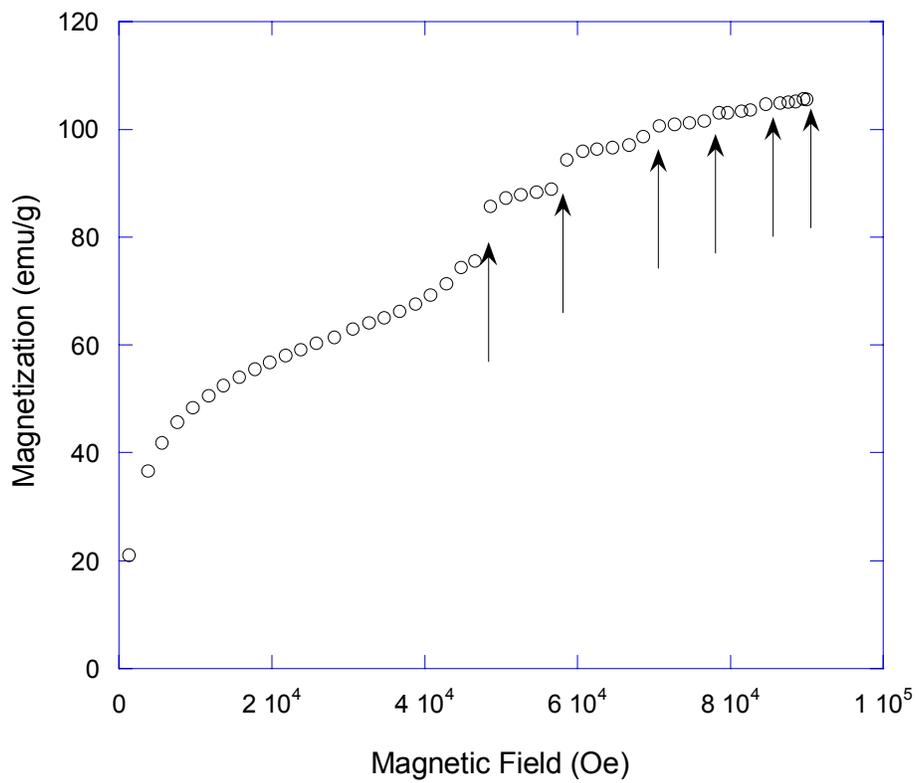

Figure 13d

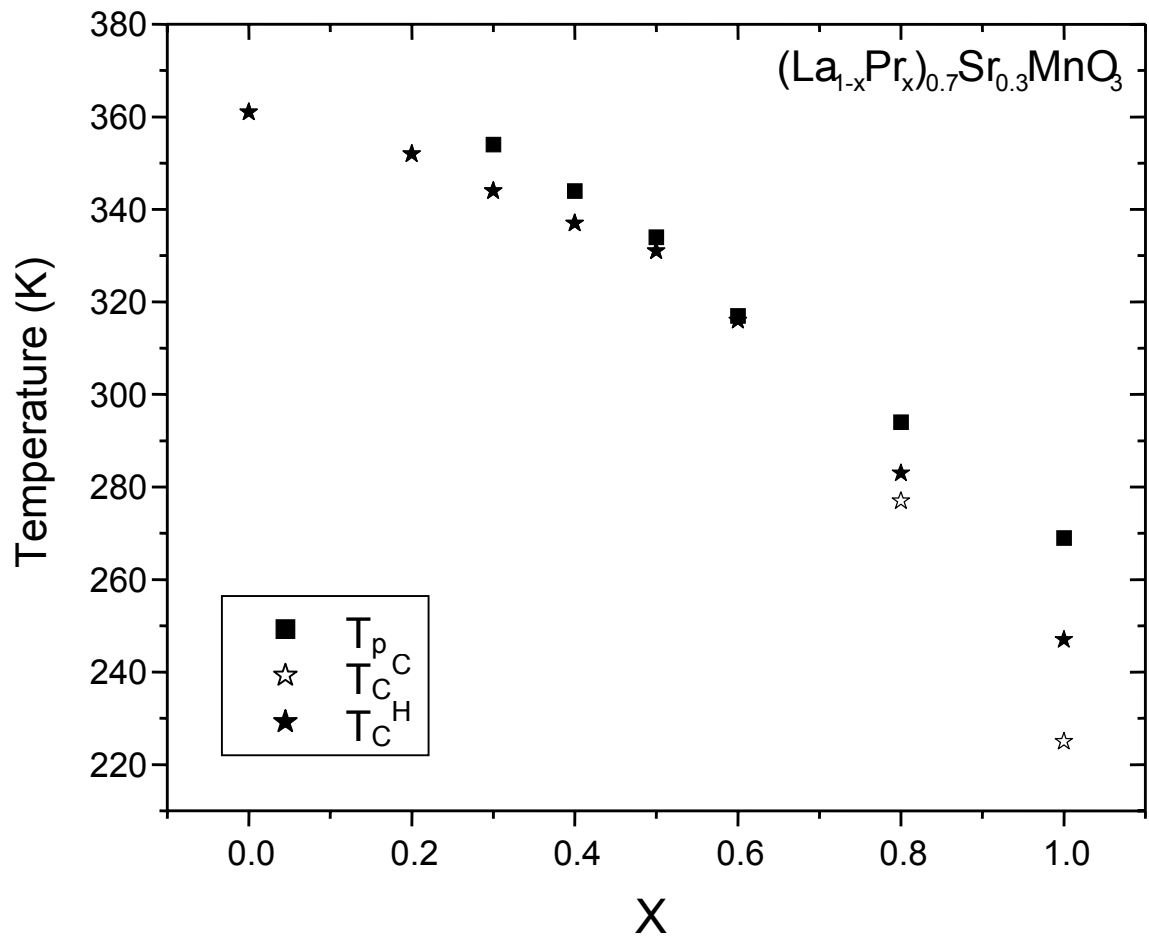

Figure 14a

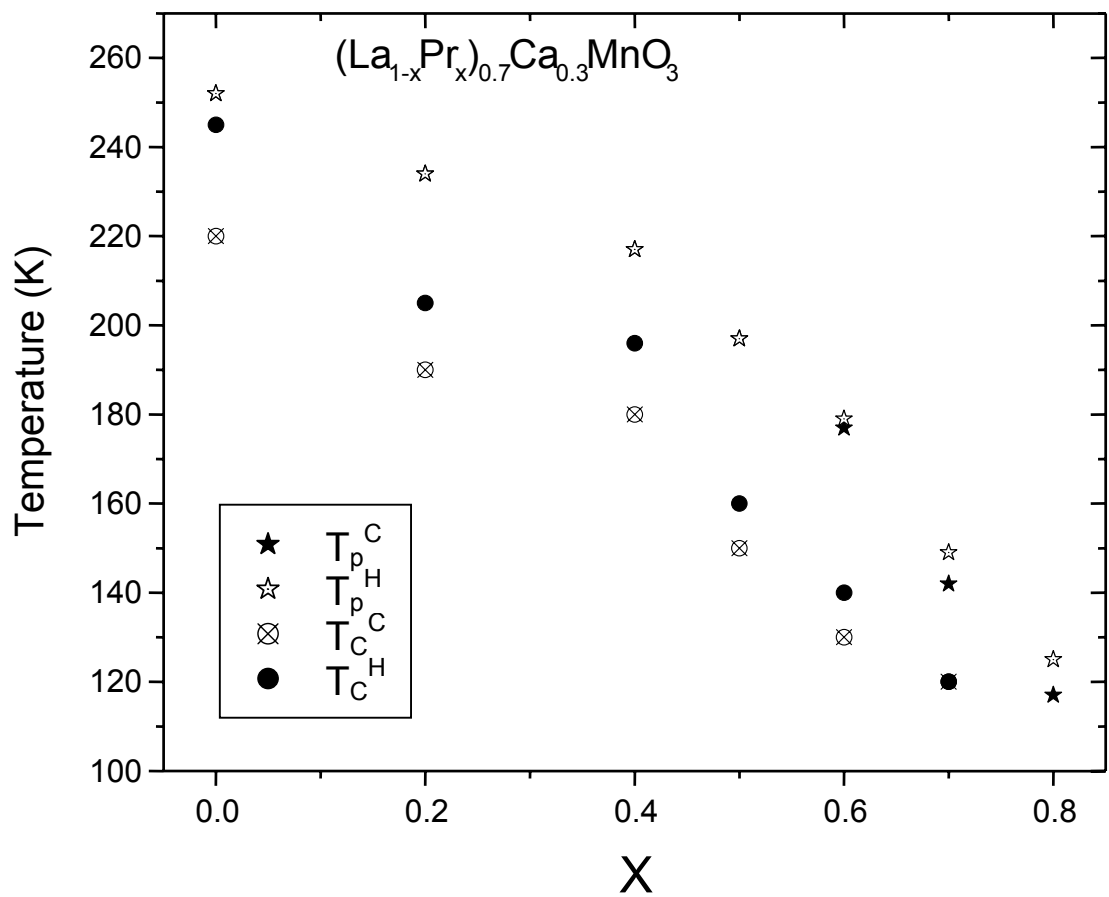

Figure 14b

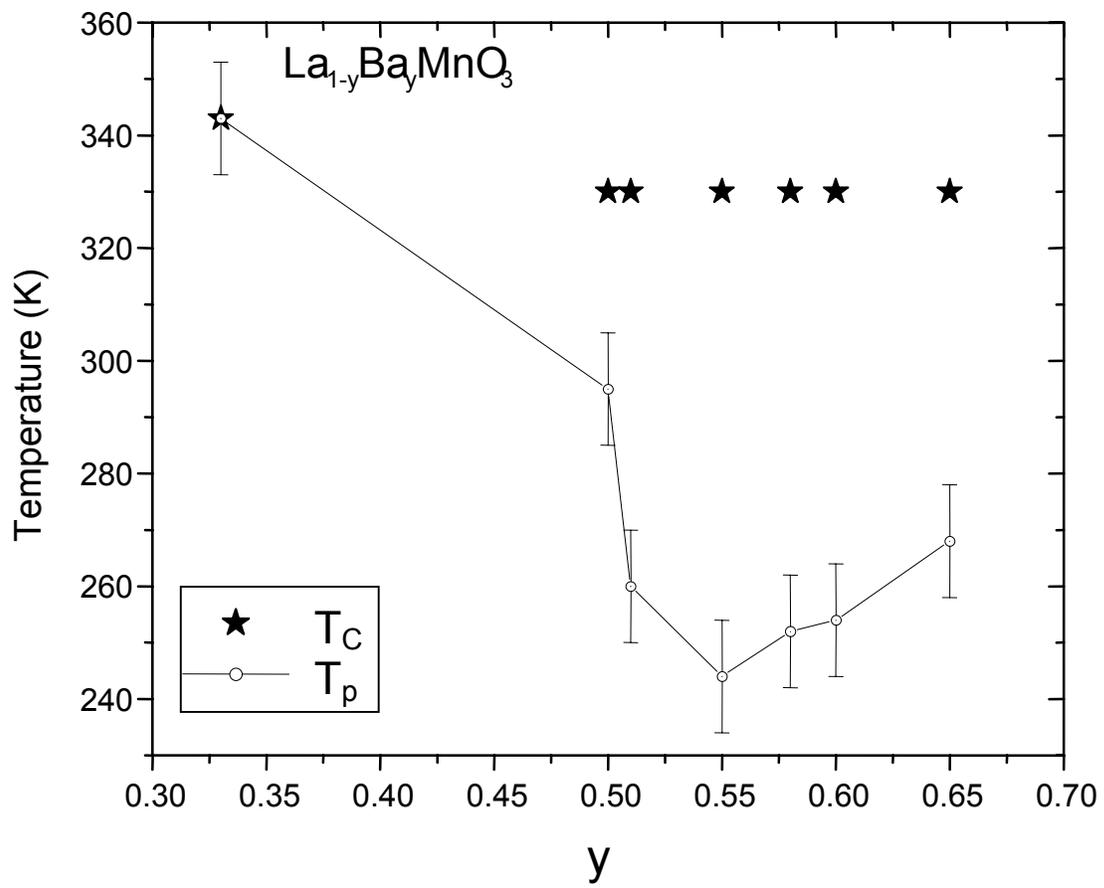

Figure 14c